\documentclass[a4paper,11pt]{article}
\usepackage{jcappub}

\usepackage{epsfig}
\usepackage{amsmath}
\usepackage{graphicx,psfrag}
\usepackage{dcolumn}
\usepackage{bm}
\usepackage{slashed}
\usepackage{xcolor}
\usepackage{maybemath}

\newcommand{\beq}{\begin{equation}}
\newcommand{\eeq}{\end{equation}}
\newcommand{\hf} {\frac{1}{2}}\color{black}

\newcommand{\rminf}{{i}}

\newcommand\fig[1]     {Fig.\,{\ref{#1}}}
\newcommand\sect[1]    {Sec.\,{\ref{#1}}}

\newcommand\app[1]     {Appendix~\ref{#1}}

\def\eq#1{(\ref{#1})}
\def\s0#1#2{\mbox{\small{$ \frac{#1}{#2} $}}}
\def\0#1#2{\frac{#1}{#2}}

\def\ord#1{{\cal O}(#1)}

\def\mr#1{{\mathrm{#1}}}

\definecolor{garrosgreen}{rgb}{0.1, 0.4, 0.1}
\definecolor{dartmouthgreen}{rgb}{0.05, 0.5, 0.06}
\definecolor{cambridgeblue}{rgb}{0.7, 0.2, 0.1}
\definecolor{cambridgeblue}{rgb}{0.1, 0.3, 1.0}
\definecolor{oxfordblue}{rgb}{0.05, 0.2, 0.7}

\definecolor{duelferred}{rgb}{0.7, 0.2, 0.1}

\usepackage{xcolor}

\definecolor{OliveGreen}{rgb}{0,0.6,0}

\title{\boldmath Renormalization--Group Running Induced Cosmic Inflation}

\author[a]{I. G. M\'ari\'an,}
\author[b]{N. Defenu,}
\author[c,d,e]{U. D. Jentschura,}
\author[f,g]{A. Trombettoni}
\author[a,d,e,1]{I. N\'andori,\note{Corresponding author.}}

\affiliation[a]{University of Debrecen, P.O.Box 105, H-4010 Debrecen, Hungary}
\affiliation[b]{Institut f\"ur Theoretische Physik, Universit\"at Heidelberg, D-69120 Heidelberg, Germany}
\affiliation[c]{Department of Physics, Missouri University of Science and Technology, \\ Rolla, Missouri 65409, USA}
\affiliation[d]{MTA--DE Particle Physics Research Group, P.O.~Box 51, H--4001 Debrecen, Hungary}
\affiliation[e]{MTA Atomki, P.O.~Box 51, H--4001 Debrecen, Hungary} 
\affiliation[f]{CNR-IOM DEMOCRITOS Simulation Center, Via Bonomea 265, I-34136 Trieste, Italy}
\affiliation[g]{SISSA and INFN, Sezione di Trieste, Via Bonomea 265, I-34136 Trieste, Italy}

\emailAdd{marian.istvan.gabor@science.unideb.hu}
\emailAdd{ndefenu@phys.ethz.ch}
\emailAdd{ulj@mst.edu}
\emailAdd{andreatr@sissa.it}
\emailAdd{nandori.istvan@science.unideb.hu}

\abstract{ 
As a contribution to a viable candidate for a standard model of cosmology, we
here show that pre-inflationary quantum fluctuations can provide a scenario for
the long-sought initial conditions for the inflaton field. Our proposal is
based on the assumption that at very high energies (higher than the energy
scale of inflation) the vacuum-expectation value (VeV) of the field is trapped
in a false vacuum and then, due to renormalization-group (RG) running, the
potential starts to flatten out toward low energy, eventually tending to a
convex one which allows the field to roll down to the true vacuum. We argue
that the proposed mechanism should apply to large classes of inflationary
potentials with multiple concave regions. Our findings favor a particle physics
origin of chaotic, large-field inflationary models as we eliminate the need for
large field fluctuations at the GUT scale. In our analysis, we provide a
specific example of such an inflationary potential, whose parameters can be
tuned to reproduce the existing cosmological data with good accuracy.
}

\begin{document}
\maketitle
\flushbottom


\section{Introduction}

Despite the efforts made to construct a standard model of cosmology  which
explains the exponentially fast
expansion~\cite{inflation,density-fluct,slow-roll} of the early Universe,
important questions remain to be answered.  According to current opinion, the
hypothetic inflaton field is assumed to slowly roll down from a potential hill
towards its minimum.  Particle physics provides us with candidates for the
inflaton field, where the equation of state required for exponential expansion
is modeled by a slowly-moving scalar condensate.  Several inflationary
scenarios exist (see e.g., Ref.~\cite{encyc}). The simplest among these is the
quadratic, large field inflationary (LFI) scalar potential, $V(\phi) = \hf m^2
\phi^2$. 

In fact, the origin and the precise mechanism of inflation is one of the most
pressing questions in our understanding of the early Universe after the Big
Bang. A first guess would involve a scalar field $\phi$ (the inflaton
field), with an action including the scalar curvature $R$,
\beq
\label{eq1}
S[\phi] = - \int d^4 x \sqrt{-g} \left[ \hf 
\nabla_\mu \phi \, \nabla^\mu \phi + V(\phi) - \frac{m^2_p}{2} R \right] \,,
\eeq
where the Planck mass $m^2_p \equiv 1/(8\pi G)$ has been used 
and $\sqrt{-g} = \sqrt{-\det g_{\mu\nu}} = a^3$ with 
$g_{\mu\nu} = \mr{diag}(-1,a^2,a^2,a^2)$.
Here, the scale factor $a = a(t)$ describes the cosmological scaling in the
Friedmann--Lema\^{\i}tre--Robertson--Walker (FLRW,
see Ref.~\cite{FLRW}) metric (in our units, the speed of light is $c
\equiv 1$ and $\hbar \equiv 1$). 

It is commonly assumed that thermal fluctuations of cosmic microwave background
radiation (CMBR) originated from quantum fluctuations of the background (the
inflaton field and the metric) during the inflationary period. Thus, details of
the self-interaction potential of the inflaton influence thermal fluctuations
of the CMBR, and can be confronted with Planck data~\cite{planck}, which
measures thermal fluctuations of the background radiation.  Of course, a
reliable model should provide agreement with observations.  For example, the
quadratic LFI model is almost excluded by recent results of the Planck
mission~\cite{planck}. By the slow-roll analysis, one can decide whether a
particular choice for an inflationary potential can serve as a viable model,
i.e., whether it agrees with Planck data.

Following Ref.~\cite{slow-roll}, one can use a modification of the original
idea of Alan Guth~\cite{inflation}, where the ground state starts from a
metastable position and then, somewhat surprisingly, rolls down very slowly to
the true minimum. This scenario, as is well known, solves a number of problems
connected with the original idea formulated in the seminal
paper~\cite{inflation}.  Yet, concerning the slow-roll scenario, one may ask
why the inflaton field should start from a particular, well-defined slow-roll
domain (this is commonly referred to as the ``initial condition problem''). 

This paper is centered around the mentioned questions and 
is organized as follows.
We provide an initial account of the interrelation of the 
renormalization-group (RG) which we propose for a 
description of the pre-inflationary period,
and the slow-roll mechanism during inflation (see Sec.~\ref{sec2}).
Details include remarks on the method of analysis
proposed by us (Sec.~\ref{sec2C}) and the suitability 
of the $\phi^6$ versus the massive sine-Gordon (MSG) 
models (Sec.~\ref{sec2D}).
We proceed to both a numerical as well as an analytic 
approach to the required RG flow equations,
and their relation to the slow-roll of the potential (Sec.~\ref{sec3}).
Conclusions are reserved for Sec.~\ref{sec4}.
Four appendices supplement the considerations.
One is for details on the relation between RG running 
and Hubble time (Appendix~\ref{app1}), another one  is 
concerned with the dependence of the slow-roll analysis on 
the particular choice of the model (Appendix~\ref{appa}).
Two more appendices deal with 
general considerations regarding 
convexity of the potential and RG running (Appendix~\ref{appc}),
and with the RG evolution of the field-independent,
constant term $V_k(0)$ (Appendix~\ref{appd}).

\section{RG and Initial Condition Problem}
\label{sec2}

\subsection{Overview}
\label{sec2A}

\begin{figure}[t!] 
\begin{center} 
\begin{minipage}{0.7\linewidth}
\begin{center} 
\includegraphics[width=0.7\linewidth]{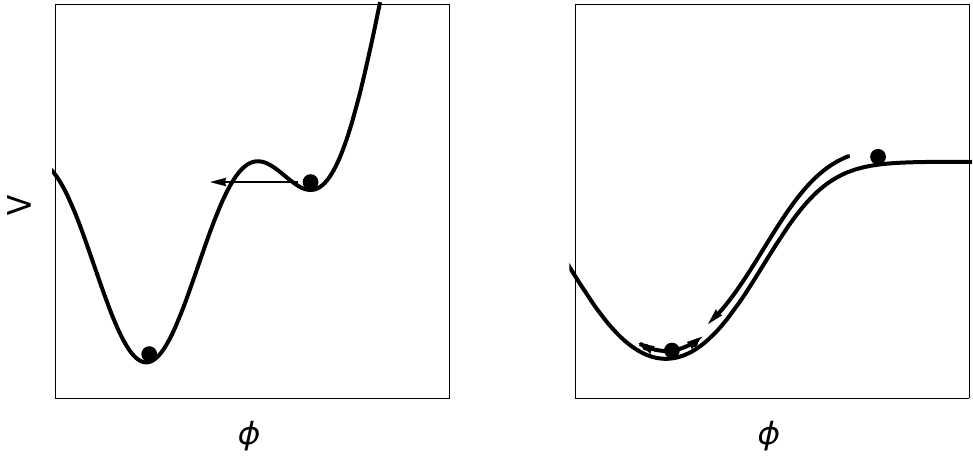}
\caption{\label{figg1} 
Inflation from false vacuum (left) and by slow-roll (right) 
where oscillations of the field produce reheating.} 
\end{center}
\end{minipage}
\end{center}
\end{figure}

The original idea of Alan Guth for inflation is based on the existence of a
relatively stable false vacuum which can be long-lived or metastable (see
Ref.~\cite{inflation}).  The system moves to the true vacuum through a bubble
nucleation caused by instanton effects via quantum tunneling; this induces
inflation. A larger energy difference between the two vacua and, alternatively,
a smaller height or width of the barrier increases the tunneling rate. Inside
bubbles, the field has its true vacuum value, and these regions of space have
lower energy, so when bubbles are nucleated, they begin to expand at nearly the
speed of light. However, this old scenario for inflation (see~\fig{figg1})
suffers from problems.

For example, it is required to heat up the Universe after the inflation period
and it is not clear how to define a proper reheating mechanism.  In order to
produce the present observable Universe, exponential expansion should continue
long enough to eliminate magnetic monopoles, but then bubbles become very rare
and in addition, they never merge.  This generates two major problems: {\em
(i)} the decay process is never complete, {\em (ii)} radiation cannot be
generated by collisions between bubble walls (which was the proposed mechanism
for radiation).

A possible solution for problems of bubble nucleations is provided by a
scenario proposed in Ref.~\cite{slow-roll}, where the ground state starts from
a metastable position and rolls down very slowly to the true minimum
(see~\fig{figg1}).  Thus, inflation  is thought to be caused by a scalar field
rolling down a potential energy hill instead of tunneling out of a false
vacuum.  The inflation is over when the hill becomes steeper. This, in
particular, solves the problem of formulating a ``graceful exit'' from the
inflation period.  

%
%

This leaves the initial condition problem to be addressed.  Chaotic
(large-field) inflation \cite{chaotic} can resolve this  issue since it is
natural to assume large field values at the energy scale of inflation, but then,
as a consequence, quantum fluctuations dominate over the classical evolution of
the inflaton field providing regions where the field moves up the potential.
This could result in eternal inflation, leading to a multiverse and the theory
could loose its predictive power. Initial conditions are even more problematic
for small-field inflation where interactions have to homogenize the Universe on
a scale larger than the horizon. Models other than cosmic inflation, like the
theory of the cyclic Universe \cite{criticism}, have also been discussed in
order to overcome the problem of eternal inflation. 

Moreover, the shape of the inflaton potential is not known except that it must
be sufficiently flat for the slow-roll  mechanism.  Therefore, the proper
treatment of quantum fluctuations is important; this may require
renormalization and, in turn, may result in scaling (running) parameters
modifying the potential itself. Indeed, the use of renormalization-group (RG)
techniques is a standard choice in the post-inflation period within
Higgs-particle-induced inflation models (for a recent overview, see
Ref.~\cite{javier_higgs_inflation}).

\subsection{RG Running and Inflationary Scenarios}
\label{sec2B}

Pursuant to the suggested relevance of RG running for the description of inflation, 
our main goal here is to discuss a mechanism which could  connect the original 
(Ref.~\cite{inflation}) and the modern (Ref.~\cite{slow-roll}) scenarios for inflation, 
or, more precisely, to investigate how the renormalization-group (RG) running of 
the inflationary potential could  imply a possible solution for the initial condition 
problem. 

During inflation itself, the slow roll-down of the VeV is commonly considered
as a classical process. 
However, our focus here is on the pre-inflationary period, 
where RG running is widely used~\cite{rg_cosmo}. Indeed, a series of papers 
are discussing the RG evolution of the Einstein equation or, more generally, 
Quantum Einstein Gravity and the Friedmann--Lama\^{\i}tre--Robertson--Walker (FLRW) 
cosmology \cite{rg_qeg} by using the ideas of Ref.~\cite{rg_cosmo}, where the 
running momentum scale $k$ of the RG approach is identified with the inverse 
of the cosmological time ($k \sim 1/t$). 
The identification of cosmological time evolution with 
RG running has been advocated, quite recently, 
in a series of papers~\cite{running_vacuum,IJMPD,PRD}, 
with a particular 
focus on the cosmological constant, which has been 
parameterized in terms of the density $\rho = \Lambda/(8 \pi G)$
[see the text before Eq.~(1) of
Ref.~\cite{PRD}, Eq.~(3) of Ref.~\cite{PRD},
and Eq.~(10) of Ref.~\cite{IJMPD}].
The model has been termed the ``running vacuum model'' (RVM).
In our model (see also Appendix~\ref{appd}),
the cosmological constant term appears as the field-independent 
term, for which we observe a perfect analogy
of our RG equation~\eqref{RGV0} with
Eq.~(3) of \cite{PRD} and Eq.~(10) of Ref.~\cite{IJMPD},
upon the identification $H \sim k$.
This observation further supports our working 
hypothesis~\cite{rg_cosmo} that $H \sim k$ can
be identified with the inverse of the cosmological time
parameter.  The RVM (see 
Refs.~\cite{running_vacuum,IJMPD,PRD})
is an extension of the $\Lambda$CDM model (cosmological constant
cold dark matter model) via a dynamical vacuum energy density.
Our work supports this hypothesis,
and it is consistent with the recent work~\cite{PRD} where
the RG running is found to be consistent 
with results derived from the bosonic (inflaton$+$axion) 
sector of an effective action which is adjusted
so that it reproduces the string scattering
amplitudes to lowest nontrivial order in an expansion in
the string Regge slope [see Eq.~(14) of Ref.~\cite{PRD}].
While in this paper, the identification 
of the RG running scale $k$ and the Hubble parameter 
is heuristically assumed, we stress, as already outlined, 
that this approach can be demonstrated
from first principles by relating the effective field theory 
under consideration with the string theory action which results in the Cosmological 
``Running Vacuum'' type model for the Universe, see \cite{running_vacuum}.
For further details see \app{app1}.

In our RG-inspired model, the VeV is assumed to be trapped in a false vacuum 
in the pre-inflationary period at very high energies, and then, due to quantum 
fluctuations, as described by the RG, the effective potential is modified releasing 
the VeV and leading to the classical inflationary evolution at the scale of inflation. 
Note that the RG evolution of the 
(self-interaction) potential of the inflaton field is a consequence
of its own scale-dependent quantum fluctuations;
the above scenario welcomes the existence of further quantum 
fluctuations below the scale of inflation, where the slow-roll process 
is otherwise considered to be ``classical''. However, the quantum
fluctuations below the scale of inflation are of less importance 
because the false vacuum has already disappeared.

The main idea behind our proposed method for RG running induced inflation is 
the fact that any scalar potential should tend to a convex one during the RG flow.
We show that the RG evolution of the potential starts from a concave potential 
at high energies, (i.e., at the Planck scale) which ensures that the VeV can be 
trapped; it tends to a less concave one at lower energies (i.e. at the GUT scale)
and releases the VeV to initiate inflation. RG running is expected to provide a 
sufficient change in the shape of the potential between the two scales. Let us note
that we show this general picture by explicit calculations with RG equations 
obtained in flat Euclidean space. One might argue that, 
nevertheless, a more reliable RG study can 
take into account for example the corresponding curved spacetime effects. 
Indeed, functional RG in an FLRW metric space has been discussed, for instance, in 
\cite{curved_RG1,curved_RG2,curved_RG3}.
The field is assumed to propagate in above mentioned 
FLRW cosmological background with the metric, 
$ds^2 = a^2(\tau) \left[-d \tau^2 + d \vec x^2  \right]$
(see also Appendix~\ref{app1}),
where $\tau$ is the conformal time and $\vec x$ denotes the comoving 
spatial coordinates [see Eq.~(1) of Ref.~\cite{curved_RG1}].
The action is defined as [see Eq.~(2) of Ref.~\cite{curved_RG2}]
\begin{eqnarray}
\label{Sphi}
S[\phi] = -\int d^4 x \sqrt{-\det g} \left[
\hf \nabla_\mu \phi \, \nabla^\mu \phi + V(\phi)  \right].
\end{eqnarray}
where $\nabla_\mu \phi \, \nabla^\mu \phi  = 
\sum_{\mu=1}^4 ( \partial \phi/\partial x_\mu )^2$ in the sense of
a four-dimensional, flat Euclidean space.
In comparison to Eq.~\eqref{eq1} above,
one drops the term $\frac12 \, m^2_p R$.
An inspection of Fig.~9 of Ref.~\cite{curved_RG3} reveals 
that RG running in the FLRW metric leads to even more enhanced 
convexification effects as compared to the flat-space
approximation used below in Eq.~\eqref{Sapprox},
which is the basis for our investigations.
Since 
convexification is an essential ingredient of our arguments, the 
proposed scenario should directly apply also to the non-flat metric 
calculations.
We identify the covariant derivative as $\nabla^\mu$ in 
Eq.~\eqref{Sphi}.

\subsection{Proposed Method} 
\label{sec2C}

Let us emphasize that the proposed RG-induced mechanism should work for any
(differentiable) inflationary potential which {\em (i)} has a concave region
and {\em (ii)} has at least one false vacuum.  If these two conditions are
fulfilled by the model, one can apply the method in three steps:
\begin{itemize}
\item One has to perform the slow-roll study of the model; this puts
constraints on the shape of the potential. As a result, the "slow-roll
potential" becomes (almost) flat in its concave region where the false vacuum
originates, and so, the VeV can roll down slowly over inflation.
\item One has to determine the RG running of the potential, i.e. the RG running of the couplings.
To generate the RG equations for the couplings a possible choice is to expand the potential into Taylor-series, i.e.,
\begin{equation}
V = V_0 + g_1 \phi + \frac{1}{2!} g_2 \phi^2 +  \frac{1}{3!} g_3 \phi^3 + \dots,
\end{equation}
which generates the couplings $g_1, g_2, g_3, ...$ specific to the particular model. 
One may expect that, given the existence of a suitable non-perturbatively renormalizable 
UV completion, the Taylor expansion is a valid method for analyzing the RG flow if an appropriate 
number of couplings are included.
\item Once the RG running of the couplings is determined, one has to choose
a model at the Planck scale, so that when it is evolved down to the
cosmological scale ($k \sim 10^{16}$ GeV), it becomes identical to the the
slow-roll potential.
\end{itemize}

\subsection{Suitability of the $\maybebm{\phi^6}$ and MSG Models} 
\label{sec2D}

The simplest model which fulfills the above 
three  conditions and has a $Z_2$ symmetry is the scalar potential 
\begin{equation}
V_{\phi^6} = 
\hf g_2 \phi^2 + \frac{1}{4!} g_4 \phi^4 + \frac{1}{6!} g_6 \phi^6  \,,
\end{equation}
with $g_2>0$, $g_4 <0$ and 
$g_6>0$. It has two false vacua and is here proved to tend to a 
convex one in the low-energy limit (see Appendix~\ref{appc}). 
The VeV is assumed to be 
trapped in (the only) false vacuum of the $\phi^6$ model at very high energies.
Near the end of inflation, the parameters of the 
model can be determined by comparison to PLANCK data 
(see the discussion in Sec.~\ref{sec3D}),
and the VeV rolls down slowly (see Appendix~\ref{appa2}).
However, the $\phi^6$ model, due to its nonrenormalizability,
cannot serve as a viable UV completion of the inflaton potential.
and therefore, we here consider the massive sine-Gordon (MSG) model as 
its simplest renormalizable UV completion (in four dimensions).

Let us first mention
that according to Ref.~\cite{planck}, one finds a certain disagreement 
of the natural inflation potential~\cite{nat_infl} with Planck results, but the 
natural (i.e., periodic) inflation model still provides better agreement
than the simplest quadratic LFI potential.
However, the combination of a polynomial in the field, and additional periodic terms,
provides for an attractive alternative scenario. For example, in
Ref.~\cite{linear_periodic}, a linear term is added to the periodic one and proposed 
as a viable inflationary potential and denoted as an axion monodromy. 
The linear term implies that the potential is not bounded from below. 

Here, our goal is to extend the periodic potential in 
such a way that:
{\em (i)} the potential has definite lower bounds,
{\em (ii)} the model has $Z_2$ symmetry, and
{\em (iii)} the model has more than one
non-degenerate minima, separated in energy by a tunable amount.  
The model represents a (nonperturbatively) renormalizable
UV completion of the $\phi^6$ model with non-degenerate minima.
The potential of the massive sine-Gordon (MSG) model,
which fulfills these requirements, reads as
\beq
\label{MSG1}
V_{{\rm MSG}}(\phi) = \hf m^2 \phi^2 + u\left[1- \cos(\beta \phi)\right].
\eeq
The full (Euclidean) action ${\cal S}$ of our theory is local, and is given by
\beq
\label{Sapprox}
{\cal S} = \int d^4 x \, 
\left[ \hf (\partial_\mu \phi)^2 + V_{{\rm MSG}}(\phi) \right] \,,
\eeq
where in comparison to Eqs.~\eqref{eq1} and~\eqref{Sphi}, we work in flat space.
In order to better understand the connection to 
the FLRW  metric which enters Eq.~\eqref{eq1},
let us observe that in a metric of the form
\begin{align}
\label{trafo1}
g_{\mu\nu} =& \; \mr{diag}(-1,a(t)^2,a(t)^2,a(t)^2) \,,
\\[0,1133ex]
\label{trafo2}
d s^2 =& \; -dt^2 + a(t) \, d \vec r^{\,2} \,,
\end{align}
with $\vec r = (x,y,z)$,
one can replace, in the action in Eq.~\eqref{Sphi},
\begin{align}
\label{trafo3}
\nabla^\mu \phi \nabla_\mu \phi \to
g^{\mu\nu} \partial_\mu \phi \partial_\nu \phi \to & \;
=
-\left( \frac{\partial\phi}{\partial t} \right)^2 + 
\frac{1}{a(t)^2} \, \left( \frac{\partial \phi}{\partial \vec r} \right)^2 \,.
\end{align}
One may stretch the spatial coordinates according to
\begin{equation}
\label{scaling}
\vec r ' = a(t) \vec r \,,
\qquad
\phi(t, \vec r) = \phi'(t, \vec r')
\end{equation}
and re-identify the primed coordinates and fields with the 
unprimed ones.
(The Jacobian of the transformation to the 
primed coordinates eliminates $\sqrt{-\det \, g} = a^3$ from the action.)
A subsequent Legendre transformation 
then leads to the Euclidean action given in Eq.~\eqref{Sapprox},
which we use for our RG analysis.
We should also clarify that,
for the considerations reported below, the scaling~\eqref{scaling} does not
matter as it does not affect the later identification of the 
running scale $k$ of our RG with a quantity that is a inversely 
related to a cosmological time parameter, 
which is all that is required (see also Appendix~\ref{app1}).

While we focus on the MSG model in the conventions of Eq.~\eqref{MSG1}, 
we anticipate that important conclusions of 
our studies for more general functional forms
(see Appendix~\ref{appa}).
The MSG model contains two adjustable parameters 
(the ratio $u/m^2$ and the frequency $\beta$) in addition to the usual 
normalization factor which can be fixed at the cosmological scale by the 
standard slow-roll analysis. 
Furthermore, we show in Appendix~\ref{appa} that an additional 
constant term does not modify the slow-roll results. 

\section{Pre--Inflation and Slow--Roll: Analytic and Numerical Approach}
\label{sec3}

\subsection{RG Flow of the MSG Model}
\label{sec3A}

If one would like to consider the RG evolution of the
parameters of a field theory (e.g. the inflationary potential discussed in the
previous section), then a natural choice is the momentum shell (functional RG)
method which is based on the RG flow equation \cite{eea_rg}. Furthermore, one
can try to connect the running momentum scale of the RG method ($k$) to the
cosmological time ($t$) as it has been done in many papers based on ideas
presented in \cite{rg_cosmo}.
In general, it is not that straightforward to consider time-dependence 
of a quantum field theory. One has to involve some extension of the theoretical
framework such as the closed-time-path (or Schwinger--Keldysh) formalism
\cite{sk,sk_new} which immediately leads to open quantum systems where typically an
external heat bath has to be introduced.  

In addition, one can assume the
presence of an external bath of particles whose energy might redshift,
providing a time-dependent background; this might induce some time dependence
which has been used in \cite{overshoot_prob} to solve the so-called overshoot
problem.  We would like to re-emphasize that we do not follow any of the above
attempts and do not assume that RG, {\em eo ipso}, necessarily induces a
time-dependence. However, it is still permissible to associate the running
momentum scale with the cosmological time, as described in detail
in Appendix~\ref{app1}.
In general, the (functional) 
RG allows to treat the effects of quantum fluctuations 
studying the scale dependence of a model via a blocking construction,
and the 
successive elimination of the degrees of freedom which lie above a running 
momentum cutoff. This procedure generates the momentum-shell 
RG (functional RG) flow equation \cite{eea_rg}
and widely discussed in
Refs.~\cite{Codello2015,Defenu2018} . 
In the local potential approximation (LPA) using 
the Litim regulator~\cite{Litim2000}, this equation
is written as
\beq
\label{opt}
\left(d- \frac{d-2}{2} \, \tilde \phi \; 
\partial_{\tilde \phi} + k\partial_k \right) \tilde V_k=
\frac{2 \alpha_d}{d} \, \frac{1}{1+\partial^2_{\tilde \phi} \tilde V_k} \,,
\eeq
where $\tilde V_k = k^{-d} V_k$ is the dimensionless scaling potential
$k$ is the running momentum, 
$\alpha_d = \Omega_d/(2(2\pi)^d)$, and
$\Omega_d = 2 \pi^{d/2}/\Gamma(d/2)$ is the $d$-dimensional solid angle.
The tilde denotes dimensionless quantities which are obtained
from the dimensionful ones via multiplication 
by an appropriate power of $k$, to take away their dimension. 
The dimensionful and dimensionless quantities are related by
the relations 
\begin{align}
\label{relations}
\tilde \beta =& \; k^{(d-2)/2} \, \beta \,,
\qquad
\tilde m = m/k \,,
\qquad
\tilde u = k^{-d} \, u \,,
\qquad
\tilde \phi =  k^{-(d-2)/2} \phi  \,.
\end{align}
It is important to note that the proposed RG treatment 
gives the same flow equations for each form of the MSG model, 
i.e., for our main model given in Eq.~\eq{MSG1},
as well as the variants discussed in Eqs.~\eq{MSG2} and~\eq{MSG3}, 
because field independent terms do not 
change the running, and the RG equation is symmetric under 
the $u \to -u$ transformation. 

The two-dimensional MSG model has 
already been considered using a momentum shell RG 
scheme in Refs.~\cite{msg_lpa,msg_beyond_lpa}. Here, we extend these 
studies to $d=4$ dimensions. 
We look for the solution of Eq.~\eq{opt} in the functional 
form of the MSG model.

We pause here to point out that the function $V_k(\phi)$,
in zeroth order in $\phi$,
is a function only of $k$; we denote it as $V_k(0)$.
Some comments and speculations on the RG evolution of the
field-independent, constant term $V_k(0)$ are presented
in Appendix~\ref{appd}. From one side, the precise form of the function
$V_k(0)$ is irrelevant for our following discussion,
in the sense that none of our results
of the main text depend on it. From the other side, we alert the reader that
if one aims at a determination of the free energy in a flat background
or of the cosmological constant in a general non-flat background,
then the problem of unambiguously determining
$V_k(0)$ has to be seriously considered.
A few remarks on this point, beyond the discussion
in Appendix~\ref{appd}, are in order.
Namely, it is known that in principle,
the nonperturbative RG method for the MSG
(see Ref.~\cite{eea_rg} and Appendix~\ref{appc}),
cannot be used in order to determine the
RG evolution of the field-independent, $k$-dependent
term $V_k(0)$, uniquely. Upon integration, such terms
amount to field-independent contributions
to the total action integral.
The reason for the emergence of the
field-independent, $k$-dependent terms lies in the
fact that properties are imposed on the
regulator [see Eq.~\eqref{regulator} and
Eqs.~(13)---(15) of Ref.~\cite{Gi2006}].
This implies that the solution of the RG equation
for the potential $V_k(\phi)$ recovers the bare action in the
UV only up to a $k$-dependent, but field-independent
term. The problem reflects the fact
that the field-independent, but renormalization-scale
dependent terms can lead to an (unphysical) 
divergence of $V_k(0)$ in the UV limit,
as explained in Sec.~2.3 of Ref.~\cite{Gi2006}. 
This unphysical behavior necessitates the introduction 
of subtraction terms if any meaningful
information is to be obtained. In Appendix~\ref{appd},
we discuss possible subtractions. 
We finally observe that the result one may obtain 
for $V_k(0)$ in the effective action
would be proportional to the free-energy 
density of the inflaton field in the flat
background limit. A systematic investigation of the conditions that could
unambiguously fix the subtraction scheme in the general case
with curved space-time is currently
lacking, to the best of our knowledge, and it would be extremely desirable.
For the purposes of the present paper, our view is to choose a
renormalization scheme by which we obtain observationally acceptable results for the RG
running of the potential. Provided
we have at least one false vacuum in the UV, we conjecture that these results, in the
IR, even become independent of the details of the particular field-theoretical model
once that, due to the RG running,
the potential starts to flatten and becomes convex. The question of whether the chosen
RG scheme is also capable of describing the RG running of the field-independent term is
of lesser importance in the
mentioned context.

Let us consider, then, the field-dependent
terms, for which the RG equation used by us requires no
subtractions or modifications.
So, substituting 
expressions given in Eqs.~\eq{MSG1}, ~\eq{MSG2} and~\eq{MSG3} into Eq.~\eq{opt},
one can match the periodic and non-periodic terms. 
The right-hand side is periodic, and therefore, the non-periodic terms are
\begin{align}
& \tilde m^2 \tilde \phi^2 
+ k\partial_k\left( \hf \tilde m^2 \tilde \phi^2 \right) -
\tilde u  \tilde \phi \sin (\tilde \beta \tilde \phi) \; \frac{d-2}{2} \tilde \beta 
\; + \;
\tilde u \tilde \phi \sin (\tilde \beta \tilde \phi) \; 
(k\partial_k \tilde \beta)=0 .
\end{align}
These also split into two separate equations, 
one for the $\tilde \beta$ and 
another for $\tilde m$, as one can show by investigating the purely
polynomial and  trigonometric terms separately.
Thus, from the non-periodic part, one immediately gets the flow 
equations for $\tilde \beta$ and for $\tilde m$,
\begin{subequations}
\label{RGsol}
\begin{align}
k\partial_k \tilde \beta_k &= \frac{d-2}{2} \tilde \beta_k  
\hskip 0.5cm \to \hskip 0.5cm
\tilde \beta_k = \beta \, k^\frac{d-2}{2}  \,,
\\
k\partial_k \tilde m_k^2 &= -2 \tilde m_k^2 
\hskip 0.5cm \to \hskip 0.5cm
\tilde m_k^2 =m^2 \, k^{-2} \,.
\end{align}
\end{subequations}
These solutions imply that according to the lowest-order RG study of the MSG
potential, the corresponding dimensionful quantities ($m,\beta$) remain
unchanged during the RG flow.  Therefore, only the dimensionful Fourier
amplitude $u$ changes when the cutoff scale $k$ is decreased from its
(high-energy) pre-inflationary value $\Lambda$ towards the (low-energy)
inflationary stage, describing the transition toward a convex potential.

The flow equation for $u$ can be obtained from the remaining periodic terms,
\beq
\label{kpku}
- \cos (\tilde \beta \tilde \phi) (d+k\partial_k)  \tilde u 
= \frac{2 \alpha_d}{d} \, 
\frac{1}{ (1+\tilde m^2 
+ \tilde u \tilde \beta^2 \cos (\tilde \beta \tilde \phi))} .
\eeq
Although the right-hand side of \eq{kpku} contains 
higher Fourier modes, here we focus on the single-mode 
approximation, which means that after the Fourier 
expansion of both sides of \eq{kpku}, only the single cosine 
terms are kept and compared. The result of this 
procedure is the flow equation for the Fourier amplitude 
$\tilde u$ which reads as
\beq
\label{flow}
(d+k\partial_k) \tilde u_k = -\frac{4 \alpha_d }{d}
\frac{1}{ \tilde \beta_k^2 \tilde u_k}
\left( 1-\sqrt{\frac{\left(1+\tilde m_k^2\right)^2}{\left(1+\tilde m_k^2\right)^2
- \tilde \beta_k^4 \tilde u_k^2}} \right) \,.
\eeq
Solving these partial differential equations for the couplings simultaneously,
one determines how the potential of the MSG model depends on the running scale.

It is useful to rewrite Eq.~\eq{flow} in terms of dimensionful quantities,
with the result
\beq
\label{flow_dimful}
k\partial_k u_k = -\frac{4 \alpha_d }{d}
\frac{k^{d+2}}{ \beta^2 u_k}
\left(1-\sqrt{\frac{\left(k^2+ m^2\right)^2}{\left(k^2+m^2\right)^2
- \beta^4 u_k^2}} \right) \,.
\eeq
While $u_k$ is running, the
two other dimensionful parameters ($\beta, m$) are constant over the flow,
according to Eq.~\eqref{RGsol}.
Let us linearize Eq.~\eq{flow_dimful} around the Gaussian fixed point assuming 
that $ \beta^2 u_k/(k^2+m^2)\ll 1$, with the result
\begin{eqnarray}
\label{flow_dimful_lin}
k\partial_k u_k &=& -\frac{4 \alpha_d }{d} \frac{k^{d+2}}{ \beta^2 u_k}
\left(1-\sqrt{\frac{1}{1 - \beta^4 u_k^2/(k^2+m^2)^2}} \right) 
\approx \frac{2 \alpha_d }{d} \frac{k^{d+2}}{ \beta^2 u_k} \,\,
\frac{\beta^4 u_k^2}{(k^2+m^2)^2} \,,
\end{eqnarray}
which can be further simplified if the (dimensionful) mass is much smaller than the
running RG scale, i.e., $m^2 \ll k^2$, in which case  the flow equation reads as
\beq
\label{flow_dimful_lin_sg}
k\partial_k u_k \approx \frac{2 \alpha_d }{d} k^{d-2} \beta^2 u_k \,.
\eeq
The latter is the linearized RG flow equation for the {\em massless} sine-Gordon model.
Its solution can be obtained analytically and for $d=4$ dimensions it has the following
form ($\alpha_{d=4} = 1/16\pi^2$)
\beq
\label{sg_sol}
u_k = u_\Lambda\exp\left[\frac{\beta^2}{64\pi^2}(k^2-\Lambda^2)\right]
\eeq
where $u_\Lambda$ is the initial value for the Fourier amplitude at the  
UV scale $\Lambda$. Within the context of our investigations,
an obvious choice is the Planck scale, 
$\Lambda = \Lambda_p = 2.4 \times 10^{18}$\,GeV. 
However, we should emphasize that our strategy here is a 
little different from what is usually employed in an RG analysis.
Namely, while the UV scale of our analysis is the Planck scale,
this is not the scale where we match the parameters of 
our model against astrophysical observations. 
Instead, we fix the value of the running Fourier amplitude $u_k$ and also the
value of
the constant dimensionful frequency $\beta$ and mass $m$ at 
the scale of inflation $k_\rminf$, e.g., at the GUT scale,
$k_{\rminf} \sim k_{\rm{GUT}} = 2 \times 10^{16}$\,GeV,
where one can match them with the slow-roll parameters ($u_0$, 
$\beta_0$, $m_0$). Once more, we emphasize that it is 
the scale of inflation $k_i$, 
i.e., the starting point of the slow-roll, not the Planck scale, 
where the parameters are matched, according to
\beq
u_{k_{\rminf}} = u_{0}, \,\,\,\, \beta=\beta_0,  \,\,\,\, m = m_0. 
\eeq
Once these values are fixed, one can easily obtain 
the initial condition $u_\Lambda$ by using for example the 
solution~\eq{sg_sol} (in its region of applicability, as discussed above)
or in more general the solutions [Eq.~\eqref{RGsol}] of the 
RG flow equations. As a final result, 
the RG scaling of the dimensionful
Fourier amplitude $u_k$ is fully determined. 

\subsection{Pre--Inflationary Period and RG Running}
\label{sec3B}

\begin{figure}[t!] 
\begin{center}
\begin{minipage}{0.7\linewidth}
\begin{center}
\includegraphics[width=0.7\linewidth]{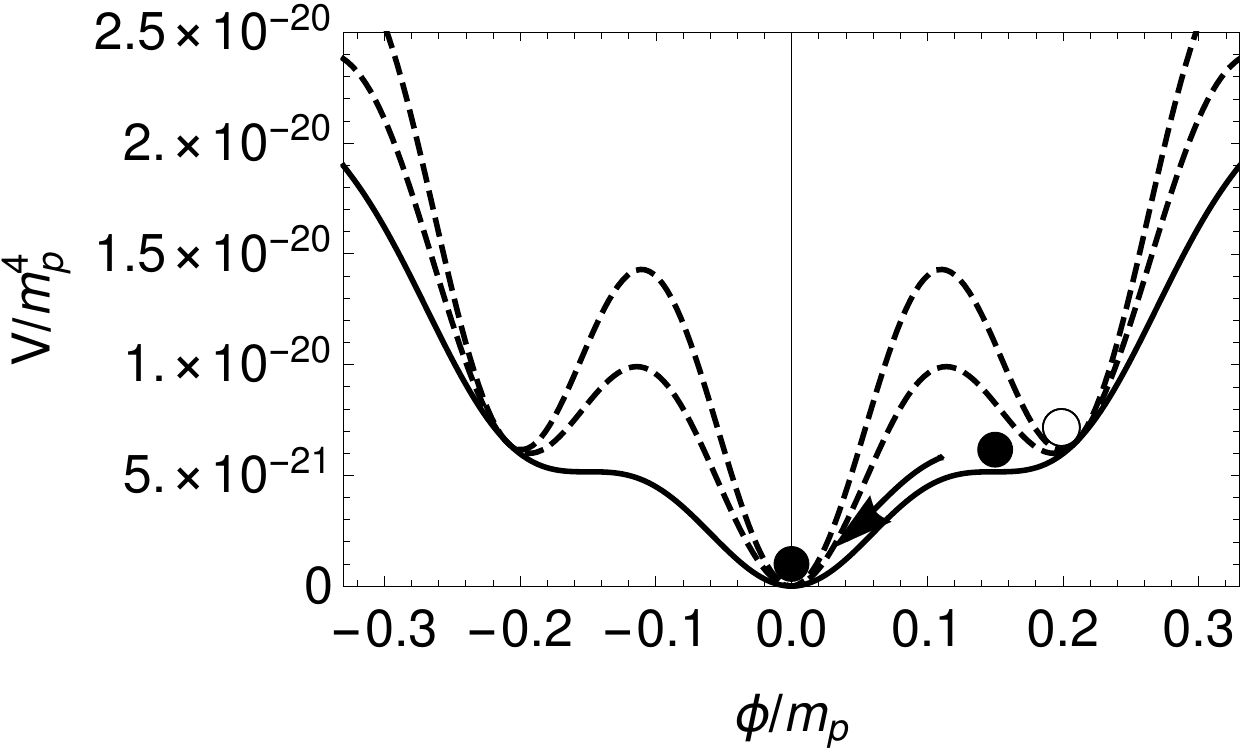}
\caption{\label{figg2} 
Potential of the MSG model \eq{MSG1} at various RG scales. 
The solid line stands for the potential at the scale of inflation 
($k_{\rminf}=2.0 \times 10^{13}$ GeV). 
The values of the potential parameters have been calculated by 
the acceptance region in \fig{figg4} with the choice $\hat \beta=30$.
Using Eq.~\eq{norm} the normalized parameters are 
$\hat m=5.4 \times 10^{-10}$, $\hat u=1.5 \times 10^{-21}$, $\hat \beta=30$ 
which yield the solid line that represents the potential over the whole inflationary 
period, where the VeV (full black circle) rolls down inducing inflation. 
The dashed lines correspond to UV values (pre-inflation), 
obtained by RG considerations using Eq.~\eq{sg_sol},  
see the discussion in \sect{sec3A}. 
}
\end{center}
\end{minipage}
\end{center}
\end{figure}

Let us now study the effects of quantum fluctuations in the 
pre--inflationary period by applying the RG method for the 
MSG model in $d=4$ dimensions. 
The quadratic LFI 
potential has no RG evolution at all, i.e., the dimensionful mass 
remains unchanged, so that the dimensionless mass has a trivial RG 
scaling. Instead, as discussed, the MSG model has a non-trivial 
RG scaling because of the periodic term which evolves under 
RG transformation; yet, the 
dimensionful mass term again has no RG evolution,
as implied by Eq.~\eqref{opt}. 
Let us now use the results of the RG analysis in order to 
discuss a possible mechanism for inflation. For the sake of 
simplicity, we consider MSG potentials with parameters fixed 
by the slow-roll study.
Suppose that at very high energy scales 
(deep UV region) in the pre-inflationary period,
the VeV is situated in the second minimum of 
the MSG potential [see \fig{figg2} for the model given in Eq.~\eq{MSG1} 
and Appendix~\ref{appa} for the other two variants,
given in Eqs.~\eqref{MSG2} and~\eqref{MSG3}].

%
%

RG evolution forces the potential to tend a convex one valid for any potential,
as shown in Appendix~\ref{appc},
and so the MSG potential becomes shallow 
at the scale of inflation. At this stage, the VeV starts to roll 
down towards the real minimum, inducing inflation. 
We find that the inclusion of quantum fluctuations 
leads to a strong renormalization of the inflationary potential, 
eventually leading to the 
disappearance of the false vacua trapping the 
VeV expectation in the pre-inflationary period, thereby inducing inflation.

\subsection{Inflationary Period and Slow Roll} 
\label{sec3C}

Let us now discuss the picture emerging for the inflationary period using 
the MSG model.  
In order to follow the conventions of inflationary cosmology 
in the slow-roll study we use reduced Planck units 
\begin{equation}
c \equiv \hbar \equiv 1 \,,
\qquad
m_p^2= \frac{1}{8\pi G} \equiv 1 \,.
\end{equation}
Thus, dimensionless quantities of the MSG model in the framework of the
slow-roll analysis are understood as 
\begin{equation}
\label{hat_def}
\hat{u} = \frac{u_0}{m^4_p}, \qquad 
\hat{\beta} = \beta_0 \, m_p, \qquad
\hat{m} = \frac{m_0}{m_p} \,, \qquad
\hat\phi = \frac{\phi}{m_p} \,,
\end{equation}
where $u_0$,  $\beta_0$ and $m_0$ are the dimensionful couplings
fixed at the scale of inflation $k_{\rminf}$, 
i.e., at a scale commensurate with or even below 
the GUT scale, and not at the Planck scale. 
In passing, we note that $m_p = 2.4 \times 10^{18}$ GeV,
while the GUT scale is at $k_{\rm GUT} = 2\times 10^{16} \text{GeV}$, so that
\beq 
\label{ratio}
\frac{k_{\rm GUT} }{m_p} \approx \frac{1}{120} \,.
\eeq
Of course, in Planck units, the dimensionful and dimensionless 
parameters (which are
commonly referred to as the ``reduced'' quantities in metrology) 
assume the same numerical values.
The slow-roll analysis fixes the (dimensionless) 
parameters, and so we can easily restore the dimension later.
In Planck units, the slow-roll conditions are the following
\begin{subequations}
\label{conditions}
\begin{align}
\epsilon \equiv V'^{2}/(2V^{2}) \ll 1 \,,
\qquad
\eta \equiv & \; V''/V  \ll 1 \,,
\end{align}
\end{subequations}
which have to be fulfilled by a suitable potential for a prolonged 
exponential inflation with slow roll down. The e-fold number 
\begin{equation}
\label{efold}
N\equiv-\int_{\phi_i}^{\phi_f} d\phi  \frac{V}{V'}
\end{equation}
should be in the range 
\begin{equation}
\label{rung}
50 < N < 60 \,. 
\end{equation}
Here, $\phi_i$ and  $\phi_f$ 
are the initial and final configurations of the field, respectively. 

The power spectra of scalar (${\cal{P_S}}$) and tensor (${\cal{P_T}}$) 
fluctuations can be characterized by their scale dependence, i.e., 
${\cal{P_S}} \sim k^{n_s -1}$, where $k$ is the comoving wave 
number. Then, slow-roll parameters are encoded in expressions 
for the scalar tilt $n_s-1 \approx 2 \eta -6 \epsilon$ and for the 
tensor-to-scalar ratio $r = {\cal{P_T}}/ {\cal{P_S}} \approx 16 \epsilon$, 
which can be directly compared to CMBR data~\cite{encyc}.

\begin{figure}[t!] 
\begin{center}
\begin{minipage}{0.7\linewidth}
\begin{center}
\includegraphics[width=0.7\linewidth]{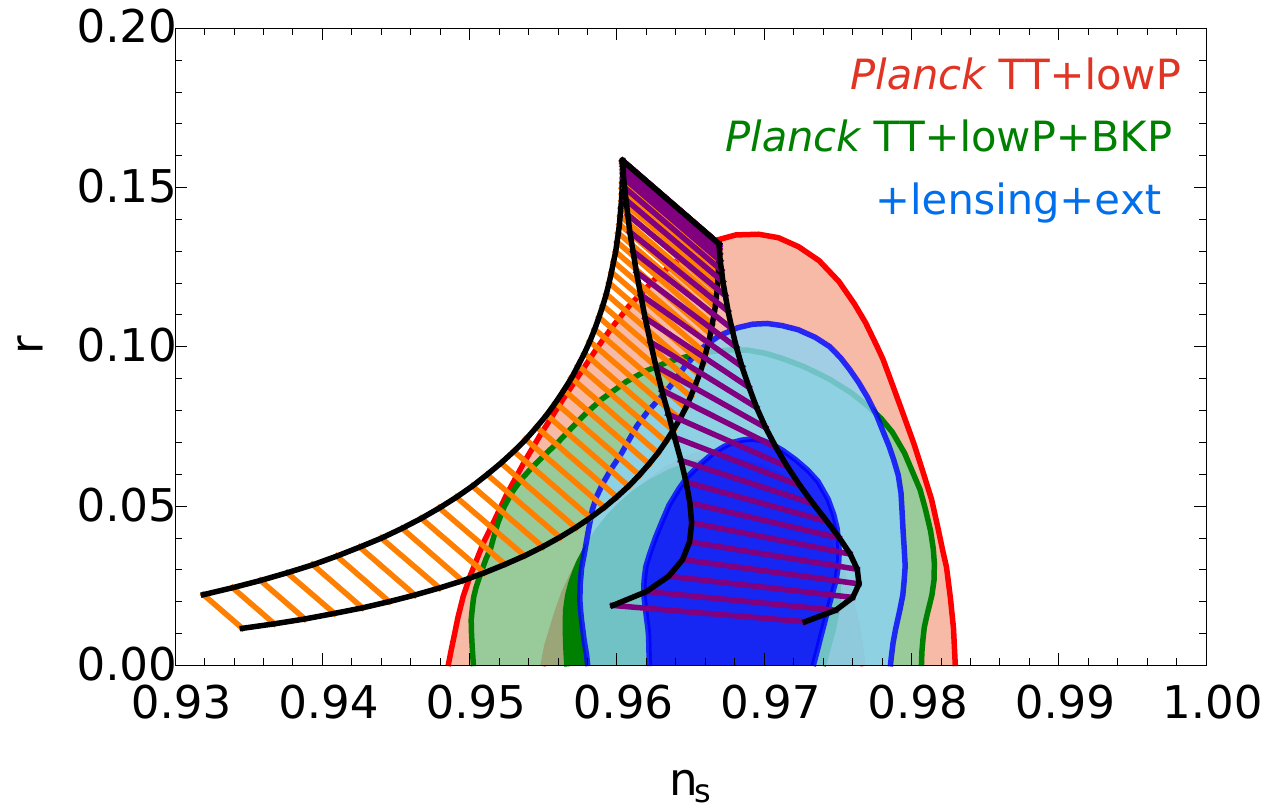}
\caption{\label{figg3} 
(Color online.)
Scalar tilt $n_s$ and tensor-to-scalar ratio $r$ derived for various frequencies 
{\em (i)} for the natural inflation model (orange line segments to the left) and 
{\em (ii)} for the MSG model given in Eq.~\eq{MSG1} with fixed ratio 
$\hat{u}/\hat{m}^2 = 1/(0.22)^2$ (purple line segments in the middle).  
As evident from the uppermost orange line segment, the two models give 
identical results for $\hat{\beta} = 0$. Different purple line 
segments result from different values of $\hat{\beta}$
(for the massive model). Both are compared to results of the Planck mission 
\cite{planck} where dark color regions stand for 95\% CL and the light color 
regions correspond to 68\% CL. The notation BKP stands for the 
combined data from BICEP2/Keck Array and Planck Collaborations.
We adopt the following labels for likelihoods: (i) Planck TT denotes the 
combination of the TT likelihood at multipoles $l \leq 30$; (ii) Planck TT+lowP 
further includes the Planck polarization data in the low-l likelihood and
"ext" is used for external data, e.g. baryon acoustic oscillations (BAO).
As evident from Eq.~\eqref{bestparam},
the best acceptance region for Planck data is reached for a value of 
$\hat{\beta} \approx 0.3$ (for fixed ratio 
$\hat{u}/\hat{m}^2 = 1/(0.22)^2$) with the massive model. 
More data (various color schemes) evidently put 
more stringent constraints on the acceptance region for $n_s$ and $r$.
The ``rungs'' of the latter 
correspond to the range~\eqref{rung} indicated for the e-fold number $N$.}
\end{center}
\end{minipage}
\end{center}
\end{figure}

\begin{figure}[t!] 
\begin{center}
\begin{minipage}{0.7\linewidth}
\begin{center}
\includegraphics[width=0.7\linewidth]{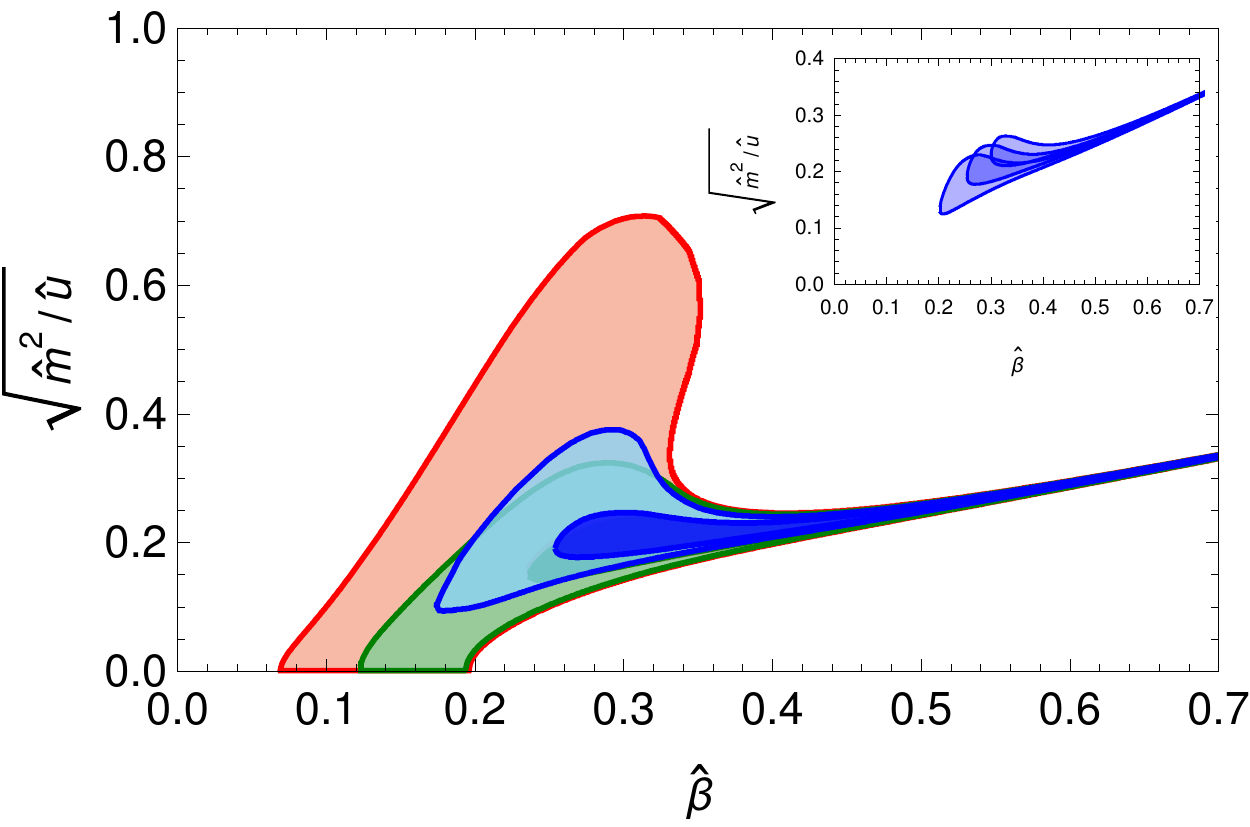}
\caption{\label{figg4} 
(Color online.)
Regions of the parameter space of the MSG model 
given in Eq.~\eq{MSG1} are indicated by 
different colors corresponding to a different level of acceptance 
for folding number $N=55$. The dark blue region gives the best fit to 
Planck data.  The inset shows the best acceptance regions for $N=50, 55, 60$.  } 
\end{center}
\end{minipage}
\end{center}
\end{figure}

We here initially focus on the first variant of the MSG model, 
given in Eq.~\eq{MSG1}. The explicit form of the potential during 
the inflationary period is reported as a solid line in \fig{figg2} (see 
also~\fig{figg7} and~\fig{figg8} and Appendix~\ref{appa}  for the 
other two variants of the MSG model).
Let us first compare results obtained from natural inflation, i.e., 
sine-Gordon and MSG models, for various values of the frequency.  
As shown in~\fig{figg3}, the MSG model provides more reliable results. 
Moreover, the ratio $\hat{u}/\hat{m}^2$ and the frequency $\hat{\beta}$ 
can be fixed by choosing the best fit to observations (see~\fig{figg3}). 
In \fig{figg4}, we indicate regions of the parameter space of the MSG 
model with different colors corresponding to different levels of acceptance.

%
%
%

%
%

For the plot in Fig.~\ref{figg2} and elsewhere, it 
is necessary to fix the scale of the ordinate axis for the potential.
This is done as follows.
Let $\phi_i$ denote the VeV of the value of the field 
$\phi$ at the onset of the slow-roll, i.e., the VeV of the potential
at the very point in the RG analysis where the ``false''
vacuum disappears and the slow-roll begins.
Writing Eq.~\eqref{MSG1} as
\begin{eqnarray}
&V_{{\rm MSG}}(\phi)& = u \, \left( \hf 
\frac{m^2}{u} \phi^2 + \left[1- \cos(\beta \phi)\right] \right) \,,
\end{eqnarray}
and in consideration of Eq.~\eqref{bestparam} below,
one immediately sees that fixing the absolute value of the 
potential at $\phi = \phi_i$ is tantamount to choosing
a particular absolute normalization for the 
value of $u = u_{k_{\rminf}}$.
Depending on the value obtained for the tensor-to-scalar ratio $r$
in the acceptance region [see also Fig.~\ref{figg3}],
the scale of inflation can be commensurate with the GUT scale, 
i.e., $k_{\rminf} \sim k_{\rm GUT} = 2 \times 10^{16} \text{GeV}$.
Here, we fix the absolute normalization factor for the potential so that 
the exact condition [see Eq.~(23) of Ref.~\cite{lyth_2}
and Eq.~(218) of Ref.~\cite{baumann}],
\beq
\label{norm}
V(\phi_i) \equiv \frac{r}{0.01} (10^{16} \, \text{GeV})^4 \,,
\eeq
is being met.  The tensor-to-scalar ratio $r$ is  
given by the slow-roll parameters, and determined 
by the $\hat\beta$-dependent acceptance region;
the scale of inflation [which is determined 
by the tensor-to-scalar ratio for a single field inflation, 
see the text above Eq.~(218) of Ref.~\cite{baumann}] is defined as 
\beq
\label{Vcond}
V(\phi_i) \equiv k_{\rminf}^4 \,,
\qquad
k_{\rminf} = \left(\frac{r}{0.01}\right)^{\frac{1}{4}} 10^{16} \,\, {\rm GeV}.
\eeq
Let us note, by way of example,
that according to \fig{figg3}, the slow-roll study of the 
MSG model with the fixed ratio $\hat u/\hat m^2 = 1/(0.22)^2$ gives
the fit parameters
$\hat\beta \approx 0.3$ and $r \approx 0.05$ 
for the acceptance region, and so, in this case, the scale 
of inflation, $k_{\rminf} = 1.5 \times 10^{16}$\,GeV, is around the GUT scale.

The theoretical predictions obtained from the MSG 
model are in an excellent agreement with observations.
Surprisingly, we can use the
results of the slow-roll study in order to fix the parameters of the 
MSG model at the scale of inflation. There 
are two regions in the parameter space of the MSG-type models where 
one finds good agreement with observations,
and we will discuss these two regions separately.

\subsection{Fit to Experimental Data for Small $\maybebm{\beta}$}
\label{sec3D}

The ratio and frequency taken from the dark blue region give 
the best fit to Planck data and the third 
parameter (the mass) can be fixed by the power 
spectrum normalization, i.e., by Eq.~\eq{norm}. Let us use this 
scenario to fix the admissible parameter regions of MSG-type models 
for small $\beta$. The best
dimensionful values for the MSG model \eq{MSG1} 
taken from the slow-roll study read can be read off
from~Figs.~\ref{figg3} and~\ref{figg4}.
In particular, the blue ``blob'' region in Fig.~\ref{figg4} can be identified
as the start point of a nearly straight line in parameter space,
along which the following condition holds,
\beq
\label{bestparam}
 \frac{\hat{u} \hat{\beta}^2}{\hat{m}^2} = 
\frac{u_0 \beta_0^2}{m_0^2}  \approx \frac{0.3^2}{0.22^2}>1 \,.
\eeq
Furthermore, the blue ``blob'' is easily located around the value 
$\hat \beta \approx 0.3$ in Fig.~\ref{figg4},
which, together with the normalization condition~\eqref{norm},
leads to the following  numerical values for the fit parameters,
\begin{subequations}
\begin{eqnarray}
\label{small_beta}
\hat{m}^{\rm{small}} &\approx& 5.92 \times 10^{-6} \, \Rightarrow \,
m_0^{\rm{small}} \approx 1.42 \times 10^{13} \, \rm{GeV}\,,  \\
\hat{u}^{\rm{small}} &\approx& 7.24 \times 10^{-10}  \, \Rightarrow \,
u_0^{\rm{small}} \approx 2.4 \times 10^{64} \, \rm{GeV}^4\,, \\
\hat{\beta}^{\rm{small}} &\approx& 0.3  \hskip 0.2cm \Rightarrow \hskip 0.2cm
\beta_0^{\rm{small}} \approx 1.25 \times 10^{-19} \, \rm{GeV}^{-1} \,.
\end{eqnarray}
\end{subequations}
For the particular choice of the fit parameters 
exemplified in~\fig{figg3},
the scale of inflation is found to be 
of the same order-of-magnitude as the GUT scale,
with the result $k_{\rminf} = 1.5 \times 10^{16}$ GeV,
as already indicated above.
Further details of the acceptance regions can immediately
be read off from Figs.~\ref{figg3} and~\ref{figg4}.
One can conclude that the slow-roll 
study for small $\beta$ provides parameters for the MSG 
model where the relation $\frac{u_0 \beta_0^2}{m_0^2} >1$
always holds [independent of the variant of 
the MSG model considered; see Eqs.~\eqref{MSG1},~\eqref{MSG2},
and~\eqref{MSG3}].

Finally, let us note that the dimensionful parameters of the MSG model determined 
by the slow-roll analysis in the small $\beta$ region are small compared to the GUT 
scale, see Eq.~\eq{small_beta}, and thus, the approximate solution~\eq{sg_sol} of the RG 
flow equation~\eq{flow_dimful} can be used to obtain the RG scaling of the potential 
and to extrapolate towards higher energy scales. 

It is easy to check that, during the RG flow, down from the Planck to the GUT scale, 
the expression $\beta \, k = \beta_0 \, k$ assumes values in the 
range
\beq
\hat\beta \, \frac{k_{\rm GUT}}{m_p} < 
(\beta_0 \, k) < \hat\beta \,.
\eeq
The change in the potential during the RG, i.e.,
during the scaling from the Planck to the GUT scale,
has to be substantial 
in order for our proposal to be physically meaningful.
From Eqs.~\eqref{sg_sol} and~\eqref{ratio}, we infer that 
\beq
\label{defchi}
\chi = \frac{u_{m_p}}{u_{k_{\rm GUT}}} = 
\exp\left[ \frac{1}{64 \pi^2} \left( \hat\beta^2 - 
\frac{\hat\beta^2}{120^2} \right) \right] \,.
\eeq
In the region
\begin{equation}
\label{70}
\hat \beta \gtrsim 30 \,,
\qquad
\chi \gtrsim 4.2 \,,
\end{equation}
the ratio $\chi$ implies a (more than) four-fold 
change in the amplitude $u$.
This constitutes the (more favorable) region of large $\hat\beta$ which we 
study in the following.
The only other way to reconsider the region of small $\hat\beta$
is to postulate an RG running over super-Planckian scales,
which however, seems somewhat problematic even for the cosmological era 
under consideration here.

\subsection{Fit to Experimental Data for Large $\maybebm{\beta}$}
\label{sec3E}

\begin{figure}[t!]
\begin{center}
\begin{minipage}{0.7\linewidth}
\begin{center}
\includegraphics[width=0.7\linewidth]{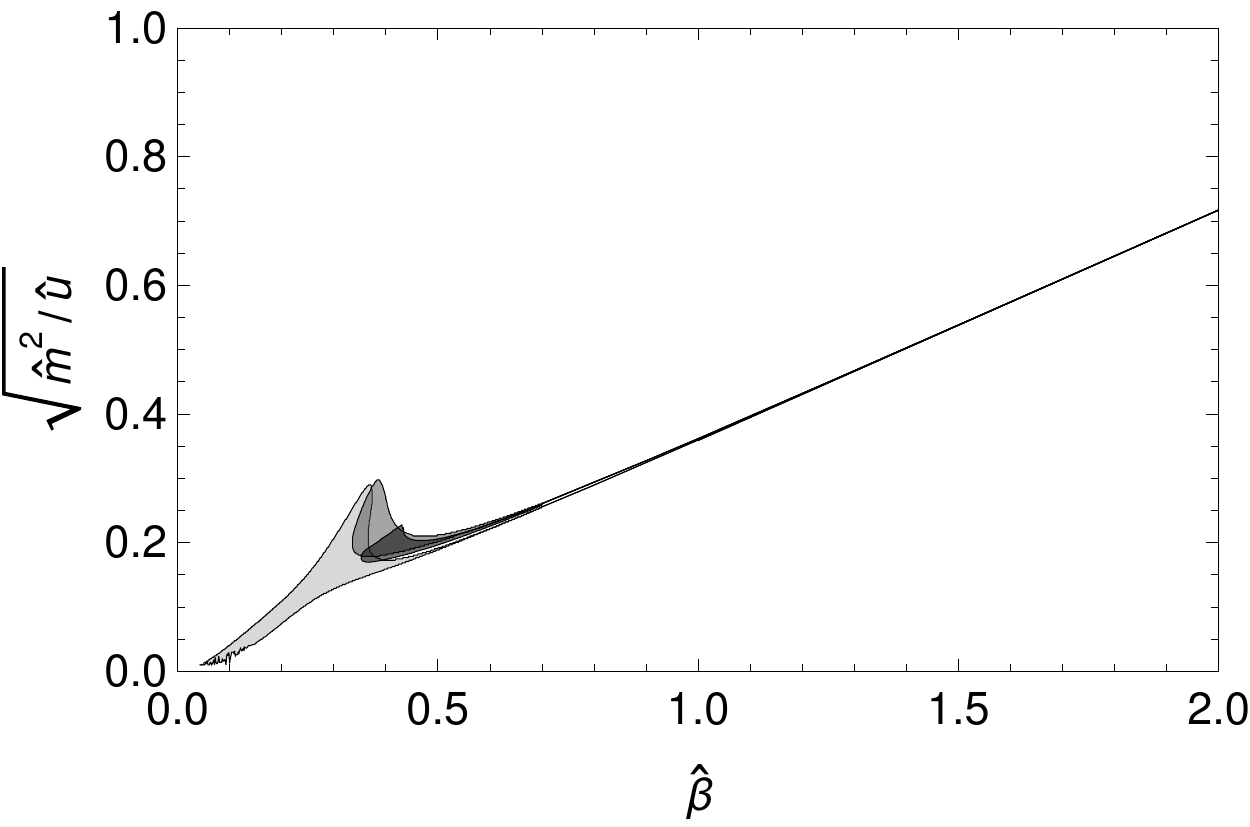}
\caption{\label{figg5}
The best acceptance regions for three variants of the MSG model
\eq{MSG2}, \eq{MSG3} and \eq{MSG4}
coincide in the limit of large $\hat\beta$.}
\end{center}
\end{minipage}
\end{center}
\end{figure}

In order to investigate the (favored) region 
of large $\hat\beta$, we have carried out numerical calculations 
for all versions of the MSG model. Those three versions [given in 
Eqs.~\eq{MSG2}, \eq{MSG3} and \eq{MSG4}], where the periodic term 
has the same sign, have absolutely equivalent behavior in the range 
$\hat{\beta} \gtrsim 1.0$, i.e., their best acceptance regions overlap 
(see \fig{figg5}). 

%
%

The first version \eq{MSG1} produces the same qualitative result, but with 
a slightly different slope. So, one can conclude that in the large frequency 
range, relevant for fixing the parameters of the model at the very beginning 
of inflation, all versions of the MSG model produce the same initial value for 
the VeV. 

\begin{figure}[t!] 
\begin{center}
\begin{minipage}{0.7\linewidth}
\begin{center}
\includegraphics[width=8.6cm]{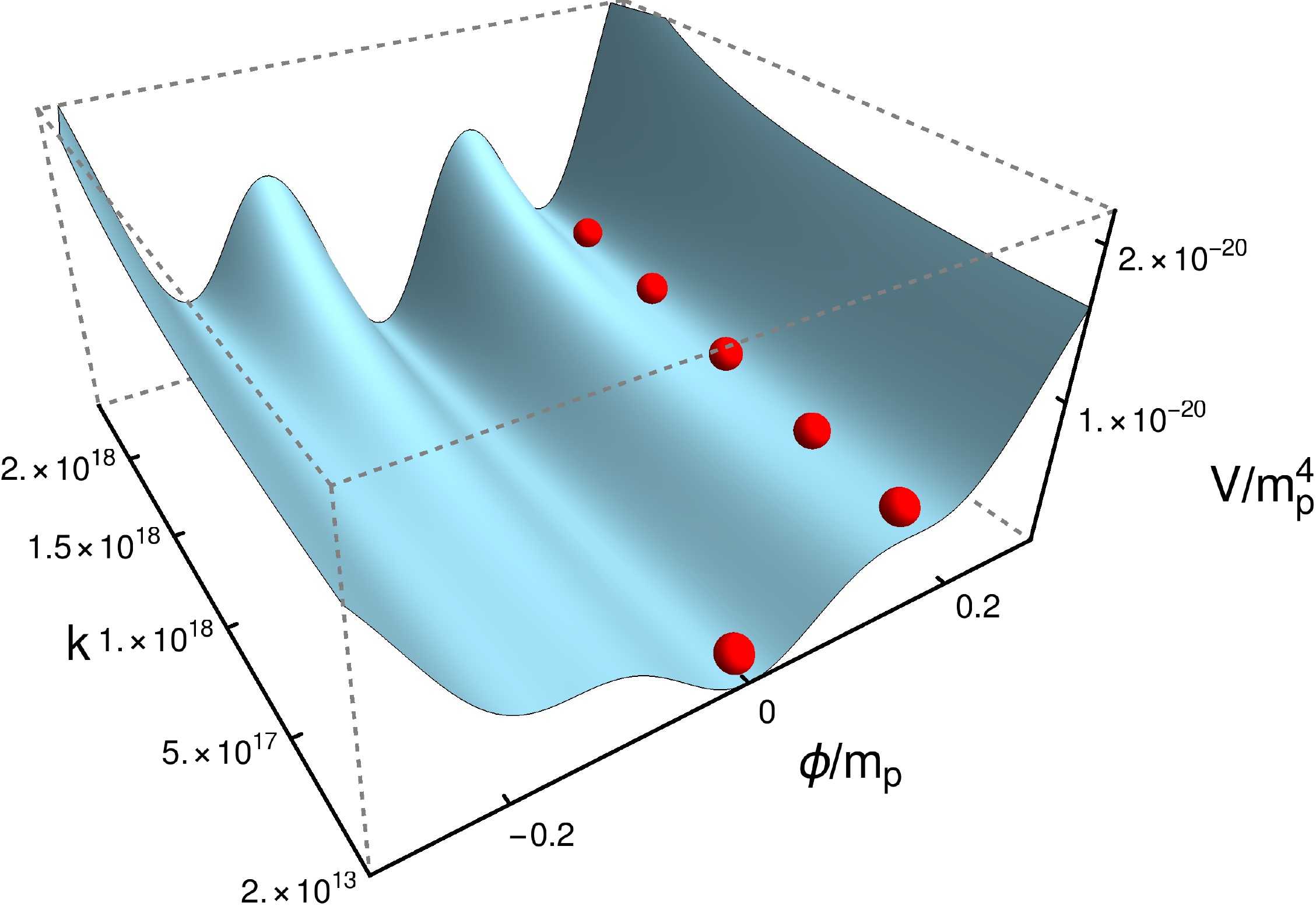}
\caption{\label{figg6}
The RG scaling of the MSG potential is shown from the Planck scale 
towards the scale of inflation for $\hat{\beta} = 30$. The MSG potential 
at $k_{\rminf} = 2.0 \times 10^{13}$ GeV is 
fixed by the slow-roll analysis. 
The red ball denotes the VeV which is trapped in a false vacuum at the 
Planck scale and due to RG running it is released at the scale of inflation
to roll down to the real ground state.}
\end{center}
\end{minipage}
\end{center}
\end{figure}

We note that the scale of inflation $k_{\rminf}$ depends on the 
particular choice of $\hat\beta$. For small values of $\hat\beta$, the inflation
occurs at the GUT scale. For large values, i.e., for $\hat{\beta} \gtrsim 1.0$ 
the scale of inflation is smaller than the GUT scale. For 
$\hat{\beta} \gtrsim 30$,
one finds $k_{\rminf} = 2.0 \times 10^{13}$\,GeV. In this particular case,
the RG running leads to a pronounced sufficient decrease in the Fourier 
amplitude $u(k = m_p)/u(k_{\rminf}) \gtrsim 4.2$. 
A visual representation of the change is given in~\fig{figg6}, where the
change in the shape of the potential against the running RG scale $k$ 
is plotted for $\hat{\beta} = 30$ [the corresponding 2D figure is \fig{figg2}].

%
%

Finally, let us add a few remarks on the 
preferred values of $\hat\beta$ that result from our analysis.
On the one hand, if one takes the view that the 
vacuum should be ``safely'' trapped in the false vacuum 
at the Planck scale, then the change in the effective 
potential from the Planck scale down to the scale of inflation
has to be substantial. Under these assumptions, a minimum choice 
would be $\hat\beta \gtrsim 30$, since otherwise there is no sufficiently large change
in the shape of the potential caused by the RG running between 
the Planck scale
and the scale of inflation; hence the RG running cannot induce inflation. 
On the other hand, if $\hat\beta \sim 300$,
then the scale of inflation exceeds the GUT scale by more than four orders of 
magnitude, which is certainly counter-intuitive.

This observation is important because the parameters of the 
model are fixed by comparison to experiment \cite{planck},
with the exception of  $\hat{\beta}$. Once $\hat{\beta}$ 
is chosen and the normalization condition~\eqref{norm} is implemented, 
all other parameters of the model are fixed. 

\section{Conclusions} 
\label{sec4}

We have demonstrated, by the  RG study of the MSG model, that the RG running of
the parameters in the pre-inflation period naturally provides us with a
flattening of the potential, i.e., it tends to a convex shape which allows the
field to roll down to the true vacuum classically (see Fig.~\ref{figg2}).  We
argue that the MSG model is a viable cosmological model, and, if  confronted
with the very recent Planck data, gives very good agreement. Our results also
show that it does not matter if we add a constant term to the MSG model, as
evident from the results obtained using variants of the MSG model.
Furthermore, we argue that the proposed RG-induced
mechanism should work for any inflationary potential with concave regions. We
merely choose the three variants of the MSG model, and the $V_{\phi^6}$ theory
to illustrate the generality of our arguments. Thus,
the very general statement holds that due to convexity, the false vacuum should
disappear at a certain lower energy scale, and the VeV which is trapped in the
false vacuum at high energies, should start to roll down towards the real
ground state when the false vacuum disappears.

However, it is important to note that in a realistic demonstration of the
change from concavity to convexity of the inflationary potential in the RG context, 
one should consider some particular GUT with a given matter content. 
This is particularly significant in the pre-inflationary epoch. We do not pursue 
such a study here, however, we here note that other fields do not change the 
RG running of the inflation potential drastically, so that the effective (scalar) 
potential in the IR should be convex.

Let us put our findings into a wider context. 
Small-field inflationary models are not in favor since initial conditions are more 
problematic for small-field inflation. Hybrid inflationary models are not supported 
or almost totally excluded by recent Planck data since they require $n_s >1$
(see Ref.~\cite{lyth_1,lyth_2}), but we know from Planck data that $n_s <1$. 
Therefore, large-field models seem to be better candidates but in general, 
they have a weakness, namely, their large initial value for the VeV which makes 
it difficult to support them by a particle-physics model. The importance of our 
method proposed here is that so-called ``chaotic'', large-field inflationary models 
could be supported by a particle-physics origin since there is no need for large field 
fluctuations at the scale of inflation. RG running induces inflation 
from a false vacuum where the VeV is trapped 
at very high energies (much higher than the scale of inflation); 
in this region, it is permissible to assume large field fluctuations.

The method of RG-running induced inflation, as we show, determines a unique
initial value as a starting point for the VeV, and so, it explains why the inflaton field 
starts in the particular slow roll domain. By a unique initial value as a 
starting point for the VeV, we mean that in the RG-induced inflationary scenario,
the VeV, i.e., the false vacuum which is the initial value for the slow-roll depends 
only on the shape of the potential. Other scenarios often depend on the random 
process of quantum fluctuations to excite the field to the initial value of the slow-roll. 
However, apart from the quantum fluctuations, the RG induced inflation has a unique 
initial value and solves this problem. In other words, no large amplitude
fluctuations are needed at the energy scale of inflation (the VeV is assumed to
be trapped in a false vacuum at very high energies).  This fact may help to
build up inflation with the same rate at various regions of space and solve the
initial-value problem.

\section*{Acknowledgments}

This work was supported by \'UNKP-17-3 New National Excellence Program of the
Ministry of Human Capacities, the National Science Foundation (Grant PHY--1710856)
and  the Deutsche Forschungsgemeinschaft (DFG, German Research Foundation) via 
Collaborative Research Centre "SFB1225" (ISOQUANT) and under Germany's Excellence 
Strategy "EXC-2181/1- 390900948" (the Heidelberg STRUCTURES Excellence Cluster).
Useful discussions with  Z.~Trocsanyi and G.~Somogyi are gratefully acknowledged. 
N.~D.~is grateful to F.~Bianchini and J.~Rubio for useful discussion in the early stages 
of this work. The CNR/MTA Italy-Hungary 2019-2021 Joint Project  
"Strongly interacting systems in confined geometries" also is gratefully acknowledged.

\appendix

\section{RG running and Hubble time}
\label{app1}
A few illustrative remarks on RG evolution and time dependence
are in order.
Here, we by no means assume that the RG, {\em eo ipso}, induces a
time-dependence. However, it is still permissible to associate the running
momentum scale with the cosmological time.  The essential idea to argue in
favor of this hypothesis has been formulated in Ref.~\cite{rg_cosmo}:
If one considers the Universe at age $t$, then it is forbidden to assume
fluctuations with a frequency higher than $1/t$. So, the mode integration,
which is the cornerstone of the functional RG method, has to be stopped at a
scale $k\approx 1/t$. For further details, one may consult Ref.~\cite{rg_cosmo}.
Indeed, a number of recent papers \cite{prd_kinvpropt} use this idea.

Let us put this statement into context. 
We first recall that the original idea to relate the 
RG to a blocking construction in real space
(which amounts to the integration of 
higher momentum shells in momentum space)
has been formulated by Kadanoff
[see Eqs.~(1)--(11) of Ref.~\cite{kadanoff}]
and Wilson [see Eq.~(17) of Ref.~\cite{wilson}].
The notion was that only some specific 
details of the microscopic theory 
should be important for the prediction of
the same critical behavior at large distance scales
(within a given ``universality class'').
In order to describe the connection
to time evolution in the Early Universe,
one imagines, e.g., a tiny crystal of a substance forming, and then, growing
fast. At first, the interactions within the 
crystal are described by the microscopic interaction, but then, as the crystal 
grows fast, the correlation functions are those of the effective, low-energy
theory.  In the same right, we can assume the Universe to grow over time, thus
realizing the Wilson blocking construction in real space, as the universe
expands. That is the idea behind the concepts 
formulated in Refs.~\cite{rg_cosmo}.
For an excellent review on the blocking construction in momentum
space and the functional RG, we refer to Ref.~\cite{polonyi}.

Indeed, invoking the concepts of homogeneity and isotropy,
it was argued in Ref.~\cite{rg_cosmo} 
that $k$ can only depend on the cosmological time, and 
that the correct identification is given by a functional relationship 
$k(t) = \xi/t$, where $\xi$ is a positive constant of
order unity. 
It would thus be assumed that the RG scale $k$ is a related to the inverse of the 
age of the Universe which is identified by the Hubble time $t \sim H^{-1}$
where $H(t)$ is the time-dependent Hubble parameter. 
However, we should stress that the precise details 
of the identification are not important for the 
fitting of the parameters of our field-theoretical model
to astrophysical observations, which will be described in the 
following. All we require is that the momentum scale $k$
be a monotonically decreasing function 
of a more general time scale, say, $\tau$, 
the latter of which constitutes some kind of cosmological 
time parameter, e.g., a comoving (particle) horizon.
An alternative  natural identification
would thus be based on the conformal time $d\tau = dt/a(t)$,
where $a(t)$ is the scale factor of the FLRW metric.  One 
finds the following expression for the comoving (particle) horizon,
\beq
\tau = \int_0^t\frac{dt'}{a(t')} = \int_0^a d \ln{a} \, \frac{1}{a H}
\eeq
[see Eq.~(36) of Ref.~\cite{baumann}].
Here, $a(t)$ is the scale 
factor of the FLRW metric.  The time-dependence of the 
comoving Hubble radius is known from the solution of the FLRW equation, 
$(a H)^{-1} = H^{-1}_0 a^{1+3w}$.
For the radiation (i.e., relativistic matter) 
dominated case, $w=1/3$ and $a(t) \sim t^{1/2}$, while for the non-relativistic matter 
dominated epoch, we have $w = 0$ and $a(t) \sim t^{2/3}$.
For the dark-energy dominated regime, one has
$w = -1$ and $a(t) \sim \exp(H t)$. The comoving 
Hubble radius is found to be an increasing parameter of 
time except for the dark-energy dominated regime
(see Table~2 of Ref.~\cite{baumann}).
For our investigations, it is important that $\tau \sim a^{1+3w}$ 
is monotonically increasing with time in the radiation-dominated regime 
($\tau \sim a \sim t^{1/2}$) and also in the matter-dominated case 
($\tau \sim a^{1/2} \sim t^{1/3}$). In our RG-supported 
picture, the precise functional relationship 
between the momentum scale $k$ of the RG and the 
cosmological time parameter $\tau$  is of lesser importance
in the  the pre-dark-matter period.
The same qualitative and, for our investigations, quantitative
picture results if one universally assumes that 
the RG scale $k$ is a monotonically decreasing function of the age of the 
very early Universe. 
Favorable and intriguing observations  
for suitable parameter regions are reported in Secs.~\ref{sec3E}.

\section{Slow--Roll Studies of Alternative Models}
\label{appa}

\subsection{Slow--Roll Study of the MSG Model}
\label{appa1}

In the framework of the MSG model, we consider the 
following variants of the model originally
proposed in Eq.~\eq{MSG1},
\begin{eqnarray}
\label{MSG1recalled}
&V_{{\rm MSG}_1}(\phi)& = \hf m^2 \phi^2 + u\left[1- \cos(\beta \phi)\right], \\ 
\label{MSG2}
&V_{{\rm MSG}_2}(\phi)& = \hf m^2 \phi^2 + u\left[\cos(\beta \phi) -1\right], \\
\label{MSG3}
&V_{{\rm MSG}_3}(\phi)& = \hf m^2 \phi^2 + u\left[\cos(\beta \phi) -1\right] - V_0 \,,
\end{eqnarray}
where the second version differs from the first only by the sign of the $u$.
The third version has an additional constant $V_0$ to keep the minima of 
the potential at zero (which can be achieved by a possibly $k$-dependent,
but field-independent subtraction term $V_0$).

In principle, one might argue that the slow-roll analysis of \eq{MSG1recalled}, with a negative 
value for the coupling $u$, should give the same result as the slow-roll analysis 
of \eq{MSG2} with the same magnitude, but opposite sign, 
of the coupling $u$, and vice versa. This conjecture could be based on the 
fact that models~\eq{MSG1recalled} and~\eq{MSG2} 
turn into each other upon the replacement $u \to -u$.
However, during our numerical 
analysis, we keep $u$ positive. Hence, there is a difference between the variants, since 
the slow-roll analysis is not symmetric under the sign change of $u$. In particular,
both $\epsilon=1/2(V'/V)^2$ and $\eta=V''/V$ depend on the sign of $u$.  An essential 
difference between the MSG variants~\eq{MSG1recalled} and~\eq{MSG2} is 
the position of the global minima. For the model~\eq{MSG1recalled},
the potential has a 
global minimum at zero, while for~\eq{MSG2},
it has two degenerate global minima, 
and a local maximum at zero. Therefore, the position of the expectation value 
of the field ($\phi_f$) after the inflation, and other results of the slow-roll analysis,
are also different for the two variants.

An essential observation is that we can, colloquially speaking, identify $u = u(k)$ 
with an ``oscillation amplitude'' of the field as it multiplies the cosine term. In some 
sense, one can thus argue that the slow-roll study of the variants of the MSG model 
depends more on the magnitude of $u$ than on its sign in the large $\beta$ limit. 
That is because, if one chooses $u$ or $-u$, and a given value of $m$ and $\beta$, 
one gets a potential whose first "false" vacuum lies approximately at the same position 
in the UV for large $\beta$ values. The "swing factor" of the potential does not 
depend on the sign of $u$, only on the magnitude (modulus) of $u$. Therefore, similar
slow-roll results are expected for the variants of the MSG model in the large $\beta$ limit
which is demonstrated in \fig{figg5}.

However, when one evolves the potential into the IR, the difference become apparent: 
The minimum of \eq{MSG1recalled} migrates to $\phi=0$, while the minimum of the 
model~\eq{MSG2} goes to a nonvanishing vacuum expectation value $\phi=\phi_0$.
We refer to Figs.~\ref{figg6},~\ref{figgAb} and~\ref{figgAc}
for three-dimensional graphics respresenting the RG evolution of the 
potentials, verifying the above, intuitive arguments.

In more general terms, it is interesting to observe that the variants \eq{MSG1recalled} 
and \eq{MSG3} are expected to give identical slow-roll results in the limit of vanishing 
mass, i.e., $m\to 0$; compare \fig{figg4}  and \fig{figgA} in the limit of vanishing mass.
This becomes evident after taking the limit $m\to 0$ in \eq{MSG3} while
the constant term, for $m\to 0$, is fixed as $V_0 = -2u$.
The variants \eq{MSG1recalled} and \eq{MSG3} then become
identical when the field variable is shifted
according to $\phi \to \phi + \pi/\beta$.

\begin{figure}[t!]
\begin{center}
\begin{minipage}{0.7\linewidth}
\begin{center}
\includegraphics[width=8.6cm]{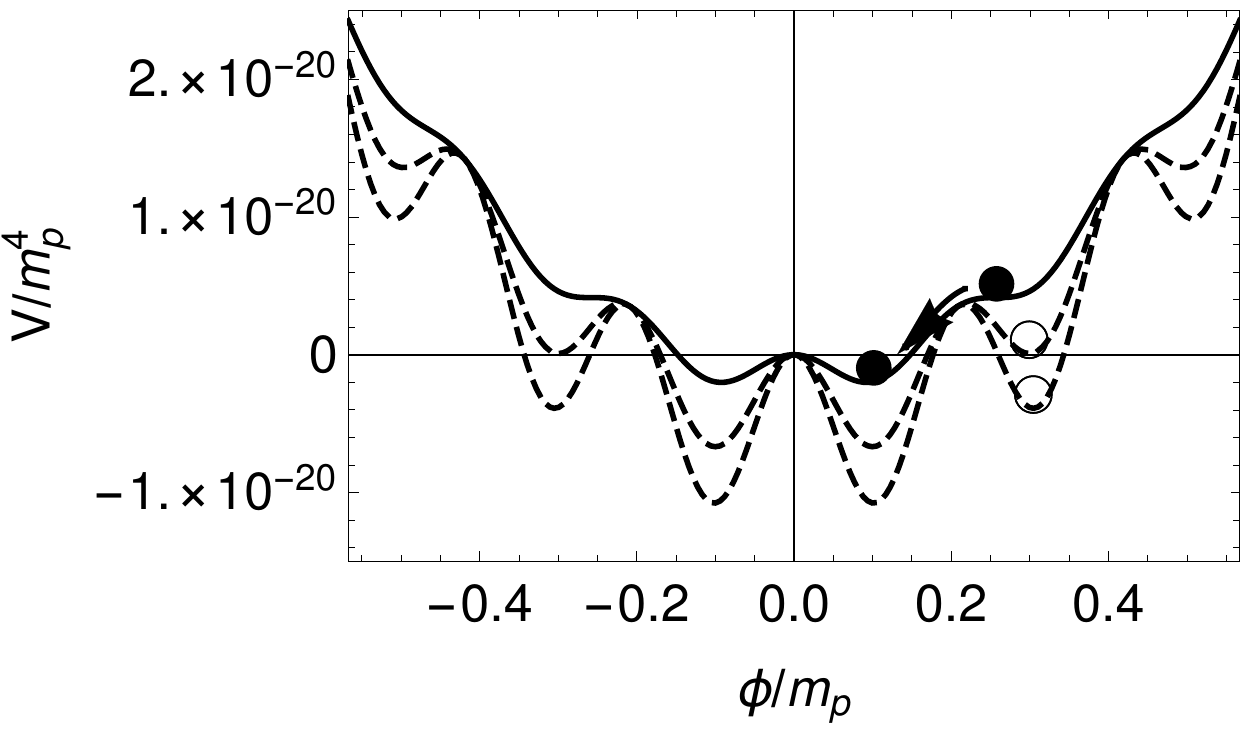}
\caption{\label{figg7}
The same as Fig.~\ref{figg2}, but for the second version of the
MSG model \eq{MSG2}.
Here the normalized parameters, corresponding to the solid line, are 
$\hat m=4.0 \times 10^{-10}$, $\hat u=1.4 \times 10^{-21}$, $\hat \beta=30$ 
which have been derived from the acceptance region of \fig{figg5} using 
the slow-roll study with $\hat \beta=30$.
}
\end{center}
\end{minipage}
\end{center}
\end{figure}

\begin{figure}[t!]
\begin{center}
\begin{minipage}{0.7\linewidth}
\begin{center}
\includegraphics[width=0.7\linewidth]{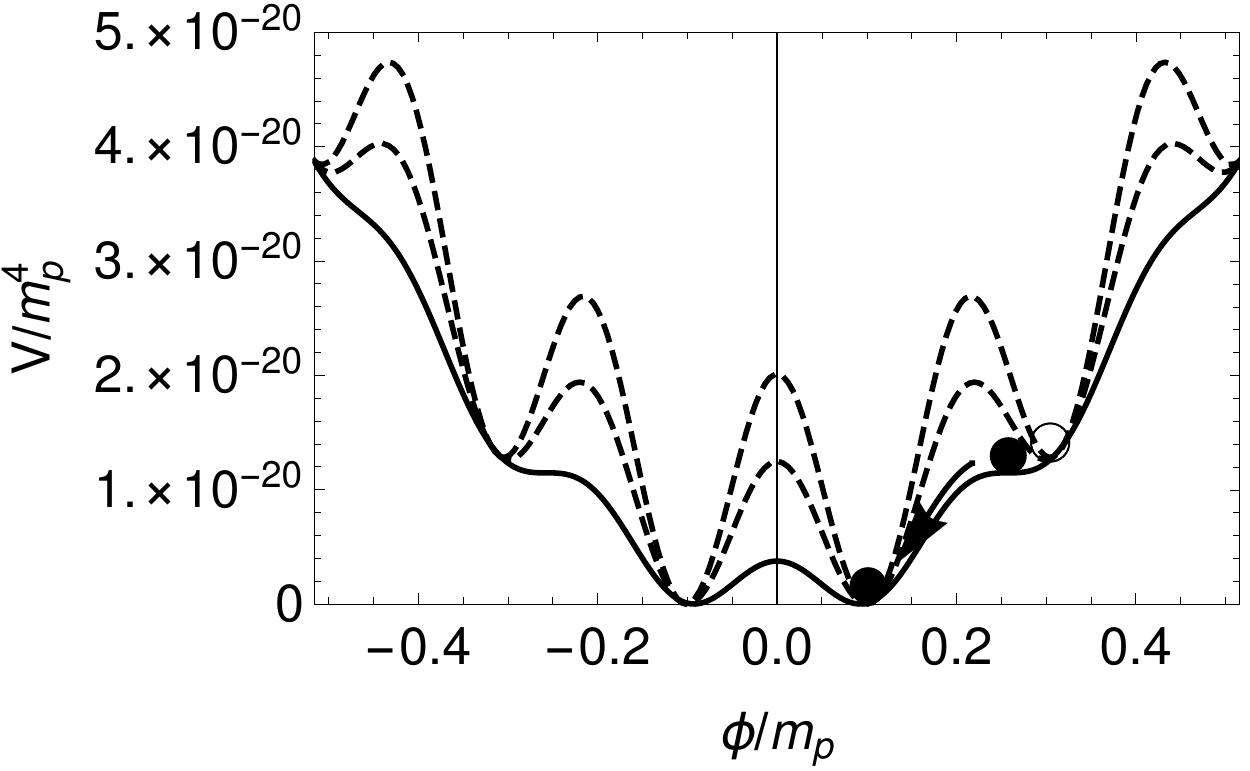}
\caption{\label{figg8}
The same as Fig.~\ref{figg2}, but for the third version of the
MSG model \eq{MSG3}.
Here the normalized parameters, corresponding to the solid line, are 
$\hat m=5.5 \times 10^{-10}$, $\hat u=2.6 \times 10^{-21}$, $\hat \beta=30$ 
which have been derived from the acceptance region of \fig{figg5} using 
the slow-roll study with $\hat \beta=30$.
Just as in Fig.~\ref{figg7} (for the second version 
of the MSG model), and in contrast to the first variant of the 
MSG model (see Fig.~\ref{figg2}), 
the third version of the MSG model predicts a nonvanishing
vacuum expectation value of the inflaton field in the IR.}
\end{center}
\end{minipage}
\end{center}
\end{figure}

Before we come to the slow-roll analysis, we 
briefly remark that the RG analysis during the 
pre-inflationary period is largely unaffected by 
the choice of the variant of the MSG model
(see Figs.~\ref{figg7} and~\ref{figg8}).

%
%

In order to have a prolonged exponential inflation with slow roll down [using reduced 
Planck units $c \equiv \hbar \equiv 1$ and $m_p^2= 1/(8\pi G) \equiv 1$], the 
conditions given in Eq.~\eqref{conditions}
must be fulfilled during inflation~\cite{slow-roll}. Indeed, the inflation stops at
$\phi_f$ if $\epsilon(\phi_{f,\epsilon})$ or $\eta (\phi_{f,\eta})$ 
reaches the value unity. Therefore, after 
substituting a potential, one can compute $\phi_f$ using the relation 
$\phi_f=\max\left(\phi_{f,\epsilon},\phi_{f,\eta}\right)$. 
The e-fold number $N$ describes how many times 
the size of the Universe got larger by an e-fold and is defined 
in Eq.~\eqref{efold}; it
is known to be in the range $50<N<60$ (see Ref.~\cite{planck}). 
For example, using dimensionless quantities, 
the parameters $\epsilon$, $\eta$ and $N$ can be expressed as follows for the first 
version of the MSG model,
\begin{align}
\epsilon=& \hf \left( \frac{\frac{\hat{u}}{\hat{m}^2} 
\hat{\beta} \sin(\hat{\beta} \hat{\phi})+ \hat{\phi}}{ \frac{\hat{u}}{\hat{m}^2} 
\left[1- \cos(\hat{\beta} \hat{\phi})\right]+ \hf \hat{\phi}^2 }  \right)^2 \,,
\\
\eta=& \frac{ \frac{\hat{u}}{\hat{m}^2} \hat{\beta}^2 
\cos(\hat{\beta} \hat{\phi})+1}{ \frac{\hat{u}}{\hat{m}^2} 
\left[1- \cos(\hat{\beta} \hat{\phi})\right]+ \hf \hat{\phi}^2 } \,,
\\
N=& -\int_{\hat{\phi}_i}^{\hat{\phi}_f} d\hat{\phi} \frac{ \frac{\hat{u}}{\hat{m}^2} 
\left[1- \cos(\hat{\beta} \hat{\phi})\right]+ \hf \hat{\phi}^2}{\frac{\hat{u}}{\hat{m}^2} 
\hat{\beta} \sin(\hat{\beta} \hat{\phi}) + \hat{\phi}} \,.
\end{align}
One finds similar results for the other two variants of the MSG model.
These quantities only depend on the equally dimensionless ratio $\hat{u}/\hat{m}^2$ 
and the dimensionless frequency $\hat{\beta}$. If the mass term is negligible compared 
to the periodic one, then one obtains back the natural inflation 
(i.e., sine-Gordon) model.
In the limit of a negligible periodic term (compared to the mass),  one obtains the 
quadratic monomial inflationary model. 

%
%

By choosing a particular value of $N$ (e.g., $N=55$), the starting point of the
inflation $\phi_i$ can be calculated using the integral~\eqref{efold} after
substituting the previously obtained $\phi_f$ for a potential. Knowing
$\phi_i$, one can compute all relevant measurable quantities. These are the
scalar tilt $n_s \approx 2\,\eta(\phi_i) -6 \,\epsilon(\phi_i)+1$ and the
tensor-to-scalar ratio $r\approx 6\epsilon(\phi_i)$. These quantities were
measured by the PLANCK mission \cite{planck} and can be directly compared to
theoretical calculations. 

\begin{figure}[t!] 
\begin{center} 
\begin{minipage}{0.7\linewidth}
\begin{center} 
\includegraphics[width=8.6cm]{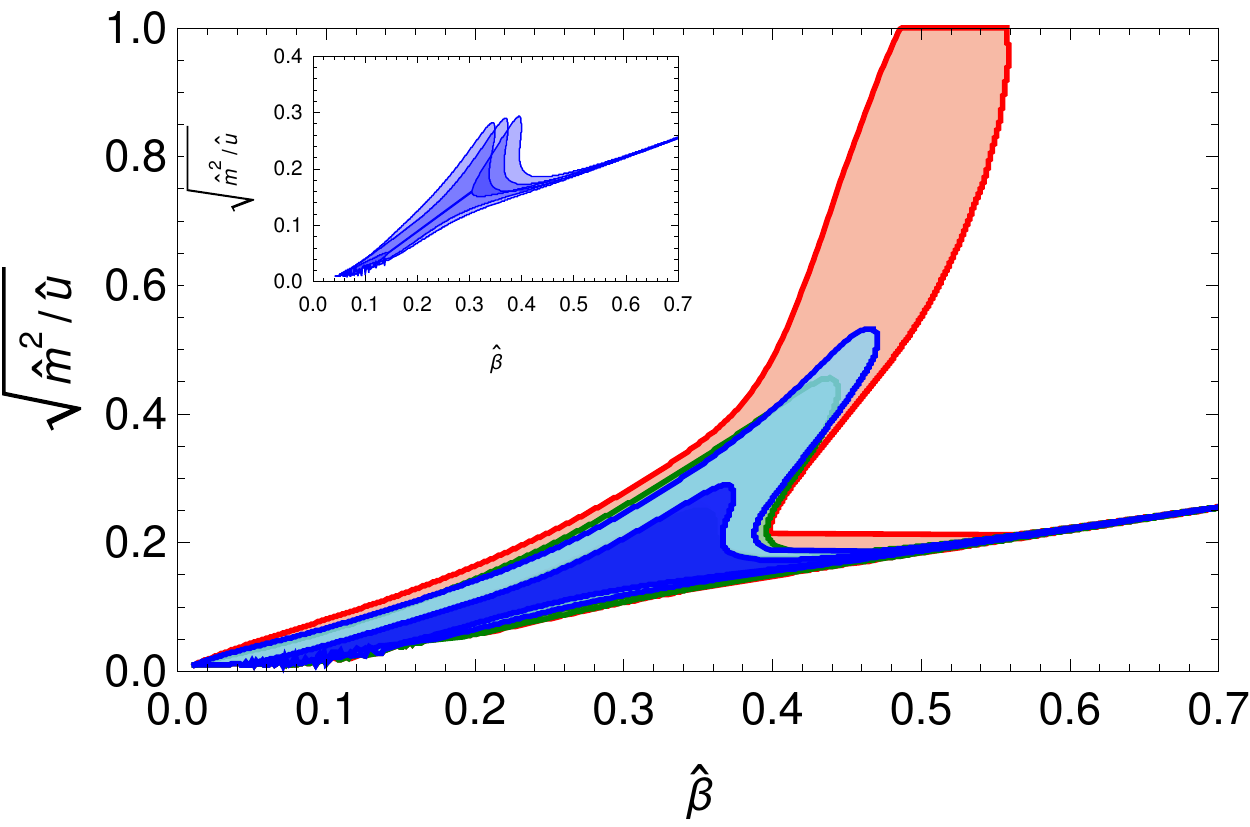}
\caption{\label{figg9}
(Color online.)
The same as Fig.~\ref{figg4},
but for the second version of the MSG model \eq{MSG2}.
Regions of the parameter space of the second 
version of the MSG model \eq{MSG2} are indicated by 
different colors corresponding to different level of acceptance 
for $N=55$, where the dark blue region 
gives the best fit to Planck data.  
The inset shows best acceptance regions for $N=50, 55, 60$.
For given data sets,
dark color regions stand for 95\% CL and the
light-colored regions correspond to 68\% CL.} 
\end{center}
\end{minipage}
\end{center}
\end{figure}

\begin{figure}[t!] 
\begin{center} 
\begin{minipage}{0.7\linewidth}
\begin{center} 
\includegraphics[width=0.7\linewidth]{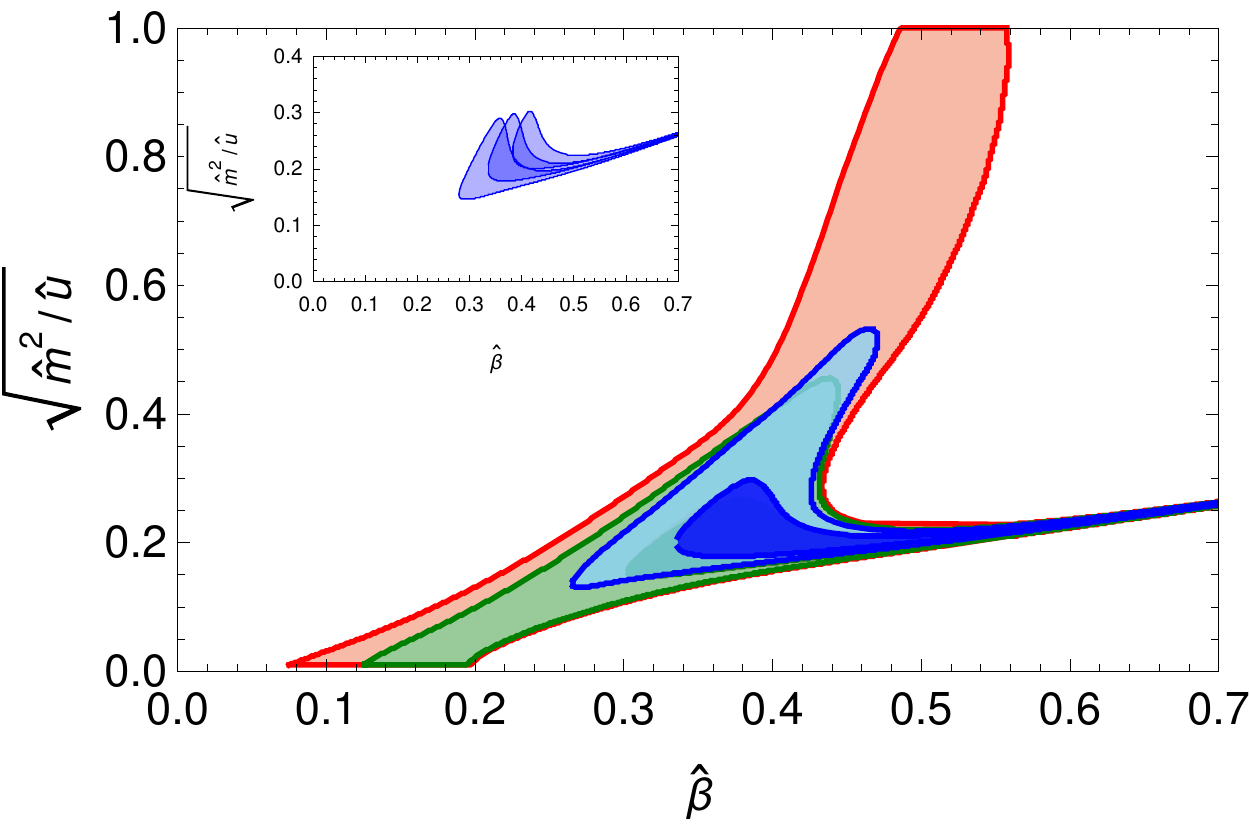}
\caption{\label{figgA}
(Color online.)
The same as Fig.~\ref{figg4} and Fig.~\ref{figg9},
but for the third version of the MSG model \eq{MSG3}.} 
\end{center}
\end{minipage}
\end{center}
\end{figure}

\begin{figure}[t!] 
\begin{center}
\begin{minipage}{0.7\linewidth}
\begin{center}
\includegraphics[width=8.6cm]{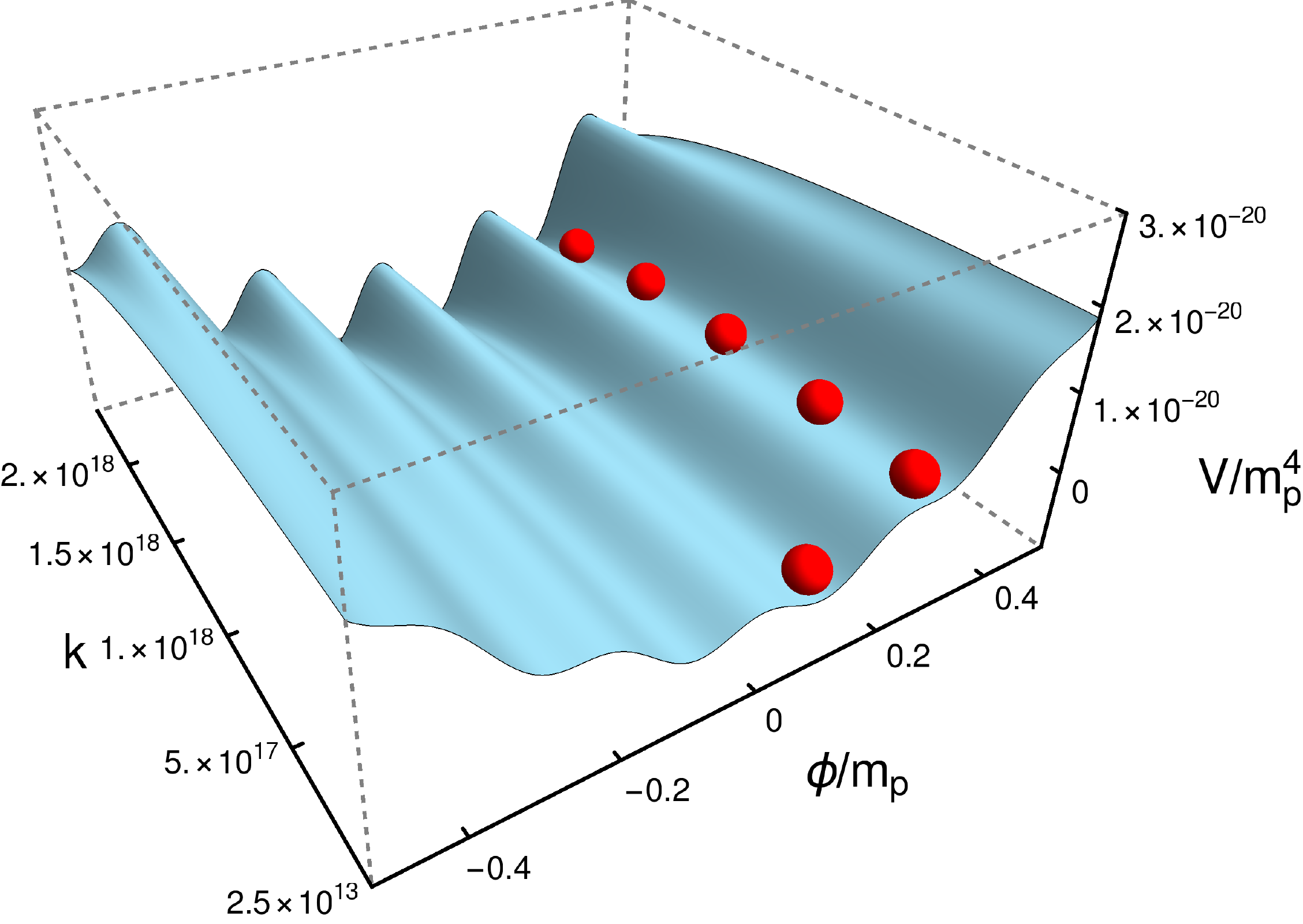}
\caption{\label{figgAb}
The same as \fig{figg6} but for the variant \eq{MSG2} with $\hat{\beta} = 30$.}
\end{center}
\end{minipage}
\end{center}
\end{figure}

\begin{figure}[t!] 
\begin{center}
\begin{minipage}{0.7\linewidth}
\begin{center}
\includegraphics[width=8.6cm]{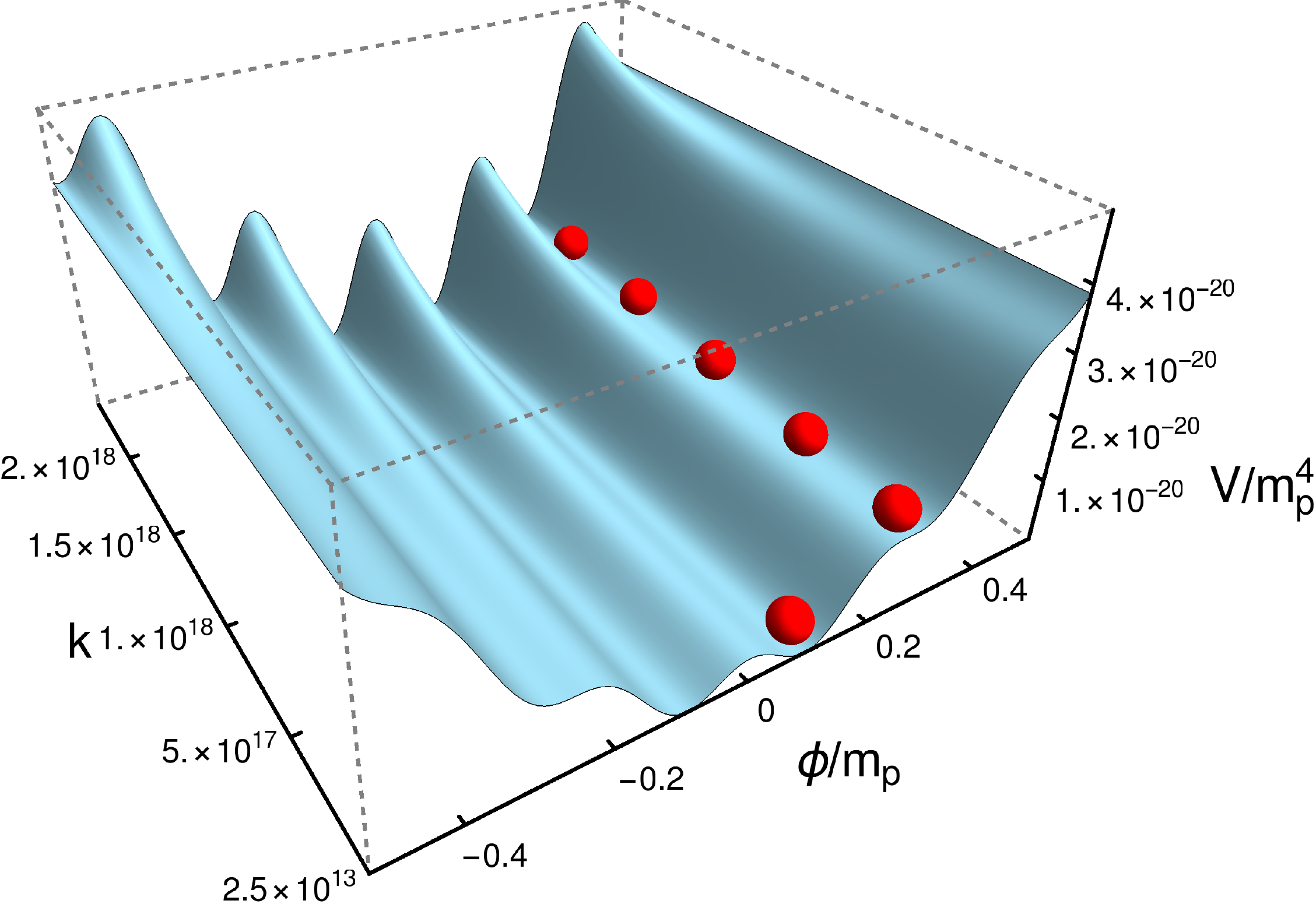}
\caption{\label{figgAc}
The same as \fig{figg6} but for the variant \eq{MSG3} with $\hat{\beta} = 30$.}.
\end{center}
\end{minipage}
\end{center}
\end{figure}

After performing the described calculations for different parameters of the
first version of the MSG potential given in Eq.~\eqref{MSG1}, we obtain 
Figs.~\ref{figg3} and~\ref{figg4},
where we indicate regions of the parameter space  with
different colors corresponding to different level of acceptance.  The same
slow-roll analysis has been performed for the last two versions of the MSG
model [Eqs.~\eq{MSG2} and \eq{MSG3}] producing very similar results (see
Figs.~\ref{figg9} and~\ref{figgA}).

%
%

%
%

In order to fix not just the ratio $\hat{u}/\hat{m}^2$ and $\hat{\beta}$ but
all three parameters of the potential, one must take into account the fact that
the magnitude of the potential when the inflation started was
$ V(\phi_i) \equiv \frac{r}{0.01} (10^{16} \, \text{GeV})^4$.  
Therefore, the best dimensionful
values for the two variants of the MSG model \eq{MSG1} and \eq{MSG2} taken from
the slow-roll study read (see \fig{figg9} and \fig{figgA}) can be determined.
For the MSG model \eq{MSG2}, one finds
\beq
\label{bestparam2}
\frac{\hat{u} \hat{\beta}^2}{\hat{m}^2} 
=  \frac{u_0 \beta_0^2}{m_0^2} \approx \frac{0.33^2}{0.17^2}>1, \hskip 0.4cm
m_0 \approx 4.93 \times 10^{-6} m_p,
\eeq
and thus $m_0 \approx 1.18 \times 10^{13}$ GeV.
For the model given in Eq.~\eqref{MSG3}, it reads as
\beq
\label{bestparam3}
\frac{\hat{u} \hat{\beta}^2}{\hat{m}^2} 
= \frac{u_0 \beta_0^2}{m_0^2} \approx \frac{0.39^2}{0.2^2}>1, \hskip 0.4cm
m_0 \approx 4.89 \times 10^{-6} m_p\,,
\eeq
where $m_0 \approx 1.176 \times 10^{13}$ GeV. Thus, the slow-roll 
study for small $\beta$ provides parameters for the MSG model where the relation 
\begin{equation}
\frac{u_0 \beta_0^2}{m_0^2} >1
\end{equation}
always holds. The best parameters of Eqs.~\eq{bestparam2} and
\eq{bestparam3} are very similar to each other in the small $\beta$ region.

Let us discuss the change in the shape of the potentials \eq{MSG2} and 
\eq{MSG3} against the running RG scale $k$. This is visually represented
in the 3D figures~\fig{figgAb} and \fig{figgAc} [the corresponding 2D figures 
are \fig{figg7} and \fig{figg8}].

%
%
%

%
%

Although in this work our focus is on the slow-roll study performed for small
frequencies, it is important to discuss very briefly the limit of 
large $\hat\beta$.
In particular, we would like to discuss how the constant (field-independent)
term influences the best acceptance results, therefore we introduce the fourth
version of the MSG model
\begin{equation}
\label{MSG4}
V_{{\rm MSG}_4}(\phi) = \hf m^2 \phi^2 + u \, \cos(\beta \phi) \,,
\end{equation}
which is obtained from Eq.~\eqref{MSG1} by the replacement 
$u\left[1- \cos(\beta \phi) \right] \to u \, \cos(\beta \phi)$. Let us consider the best 
acceptance regions for MSG type models where the periodic term has the same
sign, i.e., for the models given in Eqs.~\eq{MSG2}, \eq{MSG3} and \eq{MSG4}. 
These regions overlap for
large frequencies (see the straight (dark) line in \fig{figg5}). 
Thus, the addition constant 
term of \eq{MSG2}, \eq{MSG3} and \eq{MSG4} do not play any role in the slow-roll study. 

Finally, let us note that one of the conceivable interpretations of the MSG potential is 
that it is the UV completed potential of the Higgs, since the first two terms of its Taylor
expansion recover the usual standard model which is bi-quadratic Higgs potential.
Therefore, the $\phi$ field (in a more general model) can be a scalar doublet, but the 
potential depends only on the magnitude of this doublet and therefore, by a gauge 
transformation, it can be reduced to a single scalar field, as it is usually done when 
describing the Higgs (or, more accurately, the Brout--Englert--Higgs, or BEH) mechanism.

In conclusion, the theoretical predictions obtained from 
all types of MSG models are in an excellent agreement 
with observations in a well-defined region of parameter 
space, and therefore can be used to fix the parameters 
of the MSG models at the scale of inflation.

%
%
\begin{figure}[t!] 
\begin{center}
\begin{minipage}{0.7\linewidth}
\begin{center}
\includegraphics[width=0.7\linewidth]{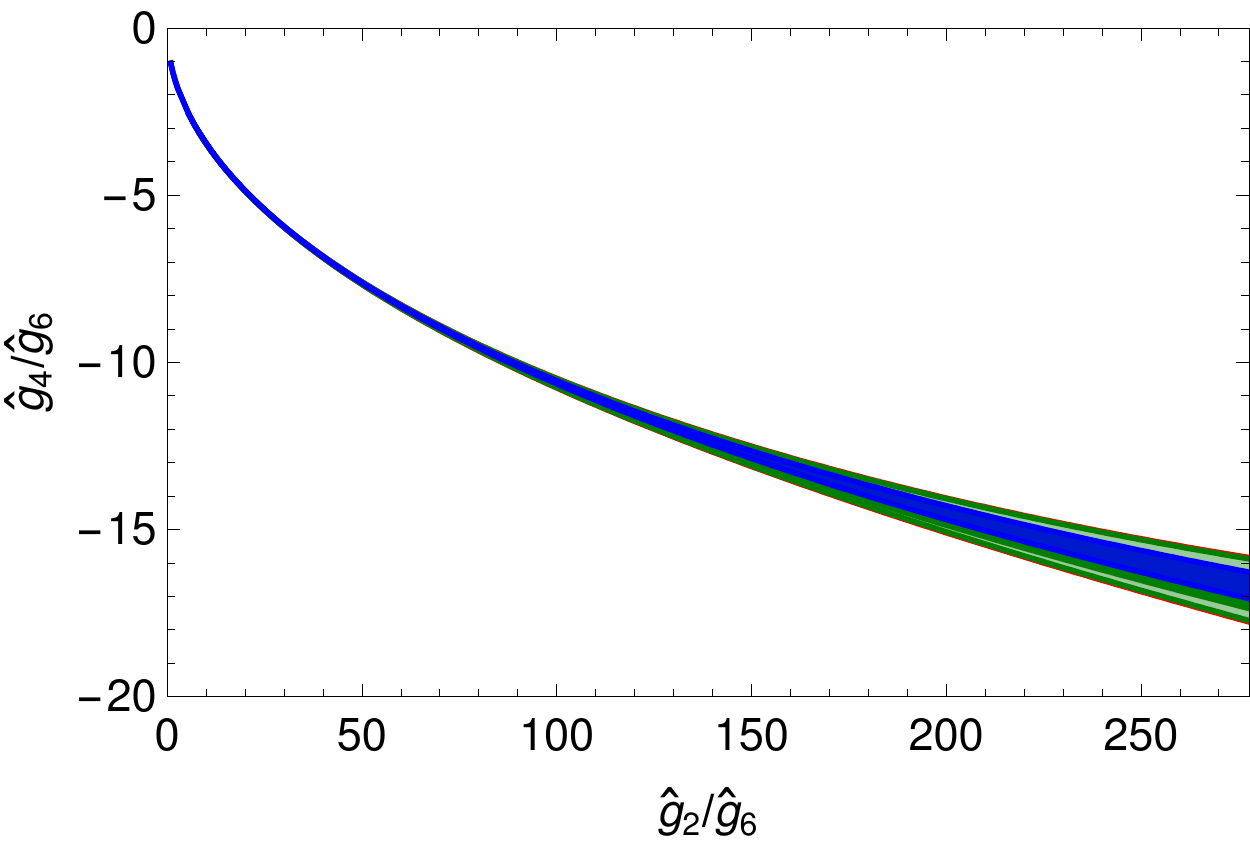}
\caption{\label{figgB}
(Color online.)
The same as Figs.~\ref{figg4},~\ref{figg9} and~\ref{figgA} given
above, but for the $\phi^6$ model \eq{phi6}. Regions of the parameter space are
indicated by different colors corresponding to different level of acceptance
for $N=55$, where the dark blue region gives the best fit to Planck data.}
\end{center} 
\end{minipage} 
\end{center} 
\end{figure} %

%
%
\begin{figure}[t!] 
\begin{center} 
\begin{minipage}{0.7\linewidth}
\begin{center}
\includegraphics[width=0.7\linewidth]{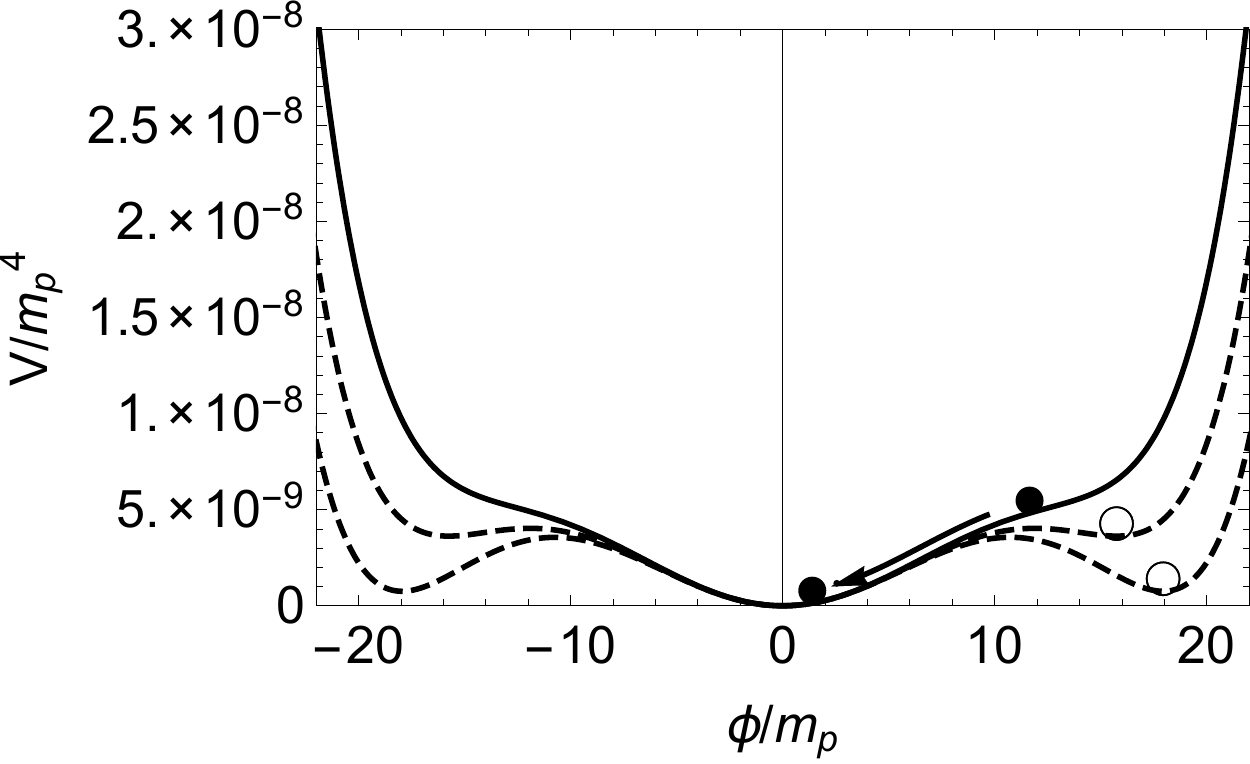}
\caption{\label{figgC} 
The shape of $V_{\phi^6}=\hf g_2 \phi^2 + \frac{1}{4!} g_4 \phi^4 +
\frac{1}{6!} g_6 \phi^6$ at various RG scales. 
The solid line stands for the potential at the scale of inflation 
($k_{\rminf}=10^{16}$ GeV), determined from the acceptance region of 
 \fig{figgB} using the slow-roll study (see \sect{appa2}) with the choice 
$\hat g_2/\hat g_6=2/6! \times 10^5$. The resulting normalized parameters  
($\hat g_2=1.4 \times 10^{-10}$, $\hat g_4=-8.4 \times 10^{-12}$, 
$\hat g_6=5.0\times 10^{-13}$)
describe the potential over the whole inflationary period, 
where the VeV (full black circle) 
rolls down inducing inflation. The dashed lines correspond to 
UV values (pre-inflation), obtained 
considering known RG equations for the couplings of the 
$\phi^6$ model, see e.g., Ref.~\cite{De2007}.
For further remarks on this point, see the discussion
aroung Eq.~\eqref{cayetano}.
} 
\end{center}
\end{minipage}
\end{center}
\end{figure}
%

\subsection{Slow--Roll Study of the $\maybebm{\phi^6}$ Model}
\label{appa2}

We also show the viability of our findings using
the slow-roll analysis of the $\phi^6$ model,
\beq
\label{phi6}
V_{\phi^6}=\hf g_2 \phi^2 + \frac{1}{4!} g_4 \phi^4 + \frac{1}{6!} g_6 \phi^6,
\eeq
which is perturbatively nonrenormalizable but 
allows for UV extensions (in the sense of the 
MSG models) which are renormalizable by a 
nonperturbative RG, as discussed at length above.
In the case of the $\phi^6$ potential, 
the key quantities, $\epsilon,\eta$ and $N$, are
described by the following expressions
\begin{align}
\epsilon=& \hf \left( \frac{
\frac{\hat g_2}{\hat g_6} \hat \phi +
\frac{1}{3!}\frac{\hat g_4}{\hat g_6} \hat \phi^3+
\frac{1}{5!} \hat \phi^5}{
\hf \frac{\hat g_2}{\hat g_6} \hat \phi^2 +
\frac{1}{4!}\frac{\hat g_4}{\hat g_6} \hat \phi^4 +
\frac{1}{6!} \hat \phi^6}  \right)^2 \,,
\\
\eta=& \frac{ 
\frac{\hat g_2}{\hat g_6} +
\frac{1}{2}\frac{\hat g_4}{\hat g_6} \hat \phi^2+
\frac{1}{4!} \hat \phi^4}{
\hf \frac{\hat g_2}{\hat g_6} \hat \phi^2 +
\frac{1}{4!}\frac{\hat g_4}{\hat g_6} \hat \phi^4 + 
\frac{1}{6!} \hat \phi^6} \,,
\\
N=& -\int_{\hat\phi_i}^{\hat\phi_f} d\hat\phi 
\frac{
\hf \frac{\hat g_2}{\hat g_6} \hat \phi^2 +
\frac{1}{4!}\frac{\hat g_4}{\hat g_6} \hat \phi^4 + 
\frac{1}{6!} \hat \phi^6 
}{
\frac{\hat g_2}{\hat g_6} \hat \phi +
\frac{1}{3!}\frac{\hat g_4}{\hat g_6} \hat \phi^3+
\frac{1}{5!} \hat \phi^5
} \,,
\end{align}
where the hat denotes dimensionless variables following the 
convention of Eq.~\eq{hat_def}, 
i.e., $\hat g_2=g_2/m_p^2$, $\hat g_4=g_4$, $\hat g_6=g_6 m_p^2$, $\hat \phi=\phi/m_p$.
These, again, do not depend on the magnitude of the potential, 
but only on the ratios of the 
couplings $\hat g_2/\hat g_6$ and $\hat g_4/ \hat g_6$.
In order to keep the potential to be bounded from below and 
have a false vacua one has to 
chose the signature $\hat g_2>0$, $\hat g_4<0$, $\hat g_6>0$ 
for the couplings.

A few remarks are in order, regarding the derivation of the 
RG equations that form the basis of Fig.~\ref{figgC}.
Namely, one might wonder how one could obtain
RG equations for the $\phi^6$ model when we have made 
the point in our paper that our main focus 
lies on the MSG model as the simplest nonperturbatively
renormalizable UV completion of the $\phi^6$ model,
precisely for the reason that the $\phi^6$ model 
is not perturbatively renormalizable in four dimensions.
To this end, we should point out that the 
RG equations used by us should be taken with a grain of salt.
Namely, they are obtained upon truncation of the 
Taylor expansion of Eqs.~(2.61) and (2.62) of 
Ref.~\cite{De2007} in powers of $\phi$, 
neglecting terms of order $\phi^8$, which would otherwise
be generated by the RG flow.
For completeness, we note the relevant RG equations
used in Fig.~\ref{figgC},
\begin{subequations}
\label{cayetano}
\begin{align}
k\partial_k \tilde g_2=& \;
-2 \tilde g_2-\frac{ \tilde g_4}{32 \pi ^2 (1+\tilde g_2)^2} \,,
\\[0.1133ex]
k\partial_k \tilde g_4=& \;
-\frac{\tilde g_2 \tilde g_6-6 \tilde g_4^2+\tilde g_6}{32 \pi ^2 (1+\tilde g_2)^3} \,,
\\[0.1133ex]
k\partial_k \tilde g_6=& \;
2 \tilde g_6+
\frac{15 \tilde g_4 \left(\tilde g_2 \tilde g_6-3 \tilde g_4^2+\tilde g_6\right)}%
{16 \pi^2 (1+\tilde g_2)^4} \,.
\end{align}
\end{subequations}
After performing the RG, as shown in Fig.~\ref{figgC},
using the same method described in the previous section, performing the
slow-roll analysis, one obtains \fig{figgB}. This shows, given the fact
that PLANCK data are sensitive only to the slow-roll of the potential (near the
end of inflation), that the $\phi^6$ model can describe the  observations made by
the PLANCK mission.  However, due to its nonrenormalizability, it cannot serve
as a viable UV completion of the inflaton potential.

\section{Convexity and RG running}
\label{appc}

Finally, let us discuss the connection between the convexity and the RG 
running of the effective action. The functional RG method is a possible choice
to determine the RG running of the parameters. 
In some sense, the functional RG equation interpolates between the classical (bare) and the 
quantum effective action. The initial condition for the RG equation is usually 
known and its determination is based on symmetry considerations. The 
RG method is used to obtain the low energy (IR) effective theory of a model 
given at the high energy (UV) momentum cutoff. However, in this work we 
have to follow a different strategy. We select between various possible 
inflationary potentials by performing their slow-roll study which fixes their 
parameters and the scale of inflation serves as an IR value for the RG 
running. Thus, as a first step, one has to explore possible 
UV completions of the 
models defined by the slow-roll study at low energies.
In some sense, we here attempt to provide a ``mini crash course''
in the functional RG suited to the problem at hand.

In order to apply the method of RG running induced inflation, the UV 
completed potential has to have a false vacuum which requires, in a 
power series {\em ansatz}, at least a term of order $\phi^6$. Moreover, in 
$d=4$ dimensions, terms of $\phi^6$ and higher powers of the field 
are nonrenormalizable perturbatively. They generates under RG flow, 
higher order terms (i.e., monomials of arbitrarily high order). This 
suggests an UV-completion of the model with a functional form that 
involves a non-terminating power series in $\phi$. These models 
{\em eo ipso} require a non-perturbative RG analysis which can be 
done by the use of the Wetterich RG equation \cite{eea_rg}, 
\begin{eqnarray}
\label{erg}
k \partial_k \Gamma_k =  \frac{\hbar}{2} \int  \frac{d^d p}{(2\pi)^d}  
\frac{k \partial_k R_k}{R_k + \Gamma^{(2)}_k}, 
\end{eqnarray}
derived for the one-component scalar field theory \cite{eea_rg}. Let us 
first discuss its connection to the effective action which has the following 
form at the 1-loop level,
\begin{eqnarray}
\label{effective_loop}
\Gamma_{\mathrm {eff}} =  S_\Lambda + \frac{\hbar}{2} \int \frac{d^d p}{(2\pi)^d} \ln\left[ S^{(2)}_\Lambda \right]  + \ord{\hbar^2},
\end{eqnarray}
where $S_\Lambda$ is the classical (bare) action. The momentum integral 
can be divergent at its upper (UV) and lower (IR) bounds. A 
momentum cutoff is a standard choice to regularize the integral; however, 
one can choose a Pauli-Villars approach by adding a momentum dependent 
mass term $\hf \int R_k(p) \phi^2$ to the bare (classical) action, and introduce 
a scale-dependent action
\begin{eqnarray}
\label{scale_effective}
\Gamma_k \equiv  S_\Lambda + \frac{\hbar}{2} \int  \frac{d^d p}{(2\pi)^d} 
\ln\left[R_k + S^{(2)}_\Lambda \right] \,,
\end{eqnarray}
which recovers the classical and the effective action (at 1-loop) in the UV and 
IR limits if the regulator function $R_k(p)$ fulfills the following requirements
[see Eqs.~(13)---(15) of Ref.~\cite{Gi2006}]
\begin{subequations}
\label{regulator}
\begin{eqnarray}
\label{regulator_prop}
R_{k\to 0}( p) &=& 0 \,, \\
\label{div}
R_{k\to \Lambda}(p) &=& \infty \,, \\
R_k(p\to 0) &>& 0 \,.
\end{eqnarray}
\end{subequations}
In a nutshell,
the first of these conditions implies that
one recovers the effective action in the limit $k \to 0$,
the second ensures that the bare action is recovered 
for $k \to \Lambda$ (within the limitations explained
in the following), and the third ensures that the 
regulator implements an IR regularization.

The divergence of the regulator for $k \to \Lambda$,
though, implies that $\Gamma_k$ only recovers the 
bare action at $k=\Lambda$ up to a field-independent,
but $k$-dependent, term.
This is also the reason why, in Eq.~(10) of Ref.~\cite{Gi2006},
as well in the text between Eqs.(15) and (16) of Ref.~\cite{Gi2006},
one writes
\begin{eqnarray}
\label{GammaK1}
\Gamma_{k\to 0} &=& \Gamma_{\mathrm {eff}} \,, \\
\label{GammaK2}
\Gamma_{k\to \Lambda} 
&=& \Gamma_\Lambda 
= S_\Lambda + {\rm const.}
\simeq S_\Lambda \,.
\end{eqnarray}
The latter identification may be part of the reason 
for the known fact (see Sec.~2.3 of Ref.~\cite{Gi2006})
that the formulation of the RG evolution of the 
constant, field-independent term requires special care
within the nonperturbative approach implied by the 
Wetterich equation.
Moreover, Eq.~\eqref{div} [when inserted into
Eq.~\eqref{scale_effective}] implies that the 
``constant term'' in Eq.~\eqref{GammaK2} actually 
is given by a divergent integral.
Let us note that the last requirement for the regulator in \eq{regulator_prop} 
is necessary to remove IR divergences in the momentum integral. As a 
second step one can differentiate Eq.~\eq{scale_effective} with respect to 
the running scale $k$
\begin{eqnarray}
\partial_k \Gamma_k = 
\frac{\hbar}{2} \int  \frac{d^d p}{(2\pi)^d}  \frac{\partial_k R_k}{R_k + S^{(2)}_\Lambda}.
\end{eqnarray}
Finally, in order to have an exact (1-loop improved) expression, the bare action
at the right hand side should be replaced by the scale-dependent one, i.e., 
$S^{(2)}_\Lambda \to \Gamma^{(2)}_k$ (and both sides multiplied by $k$) 
which results in the ``exact'' functional RG equation \eq{erg}.

Now, let us show why the effective potential should be convex \cite{pathintbook}.
The effective action is the Legendre transformation of $W[J]$,
which is  the generating 
functional for the connected Green-functions
\beq
\Gamma_{\rm{eff}}[\phi]=-W[J]+\int J \phi \, .
\eeq
If one fixes the field and the source to constant values (${\phi(x)=\phi_0}$, ${J(x)=J_0}$) 
in the space-time volume $\Omega$, then the effective action reduces to the effective 
potential as follows
\beq
\Gamma_{\rm{eff}}[\phi]=\Omega \, V_{\rm{eff}} (\phi_0) \, ,
\;\;\;
W[J]=\Omega w(J_0) \, .
\eeq
Using the relations of the Legendre transformation for the reduced functions
\beq
\phi_0=\frac{\delta w[J_0]}{\delta J_0} \, ,
\;\;\;
J_0=\frac{\delta V_{\rm{eff}}[\phi_0]}{\delta \phi_0} \, ,
\eeq
one obtains the following equation by differentiating the effective potential with 
respect to the field and the source by using the chain rule,
\beq
\label{Veffdpdj}
\left( \frac{\delta^2 V_{\rm{eff}} }{\delta \phi_0 \delta \phi_0}  \right)
\left( \frac{\delta^2 w                 }{\delta J_0    \delta J_0     }  \right)
=1.
\eeq
The second derivative of the generating functional of the connected Green 
functions with respect to the source term is the connected correlation function
which should be positive. Thus, the convexity of the effective action comes 
from Eq. \eq{Veffdpdj}
\beq
\left( \frac{\delta^2 V_{\rm{eff}} }{\delta \phi_0 \delta \phi_0}  \right) \geq 0 \,.
\eeq
On one hand, the non-perturbative RG equation \eq{erg} should recover the full 
quantum effective action in the IR ($k\to 0$) limit. On the other hand, 
we have just shown that the effective 
action or more precisely the effective potential should be convex. Therefore, if one 
performs the RG study of a possible UV-completion of the $\phi^6$ model \eq{phi6}, 
the false vacua start to vanish under the RG flow creating first a flat region of the 
potential, ideal for the slow-roll, see \fig{figgC}, which is the cornerstone of the RG 
running induced inflation method proposed in this work.

\section{RG Evolution of the Constant Term}
\label{appd}

A relatively subtle problem should be mentioned.
Namely, from Eq.~\eqref{opt}, one could in 
principle derive an RG equation
for the constant term $V_k(\phi = 0)$. 
(Re-)written for the 
dimensionful potential, the (naively obtained) RG equation
for $V_k(0)$ reads as follows ($d=4$),
\beq
\label{RGV0}
k\partial_k V_k(0) 
=
\frac{k^4}{32\pi^2} 
\frac{k^2}{k^2+\partial_{\phi}^2 V_k(0)} \,.
\eeq
In the limit $k^2 \gg \partial_{\phi}^2 V(0)$,
the solution to Eq.~\eqref{RGV0} is 
\beq
\label{largek}
V_k(0)
=
V_\Lambda(0)+\frac{k^4-\Lambda^4}{128\pi^2} \,,
\eeq
which would otherwise indicate a rampant quartic
divergence of $V_k(0)$ in the limit of large $\Lambda$,
and lead to a considerable change in
$V_k(0)$ between the Planck and the GUT scales,
possibly requiring 
fine-tuning of the model in the UV limit, i.e., at the 
Planck scale. The problem is important since 
$V_k(0)$ is associated with the cosmological
constant term.

In order to fully address the problem,
it is necessary to include a longer discussion.
We first observe that 
the functional form of
the potential given in Eq.~\eqref{MSG1}
implies that $V_k(0) = 0$ irrespective
of the numerical values of $m$, $u$, and $\beta$,
while the full potential vanishes for any field 
only if $m = 0$ and $u = 0$.
The nonperturbative RG running of the potential
could still imply the possibility of
a $k$-dependence of $V_k(0)$, which,
in the large-$k$ limit, could be deemed to be described 
by Eq.~\eqref{largek}, in which case it
becomes independent
of the numerical values of $m$, $u$, and $\beta$.
The latter observation is consistent 
with the general argument presented below,
which implies that the rampant $k^4$ behavior
is due to the emergence of the spurious terms from the 
UV properties of the regulator used in the formulation
of the nonperturbative RG equation [see Eq.~(15) of 
Ref.~\cite{Gi2006}]. Indeed, the limit of a vanishing potential ($u \to 0$, $m \to 0$) 
must be approached smoothly, and subtractions to the 
RG equation~\eqref{RGV0} are required. 

As suggested by the 
quantum mechanical calculation [see Eq.~(37) of Ref.~\cite{Gi2006}],
one is led to the following subtraction
\begin{equation}
\label{subt}
\frac{k^2}{k^2+\partial_{\phi}^2 V_k(0)} \to
\frac{k^2}{k^2+\partial_{\phi}^2 V_k(0)} - 1 \, ,
\end{equation}
which reduces the divergence of the RG evolution in the 
UV and alleviates the fine-tuning problem in the UV.
Subtractions similar to Eq.~\eqref{subt} 
are well known to produce correct results
for the free energy in quantum-mechanical models,
and in statistical mechanics. In the
present case, this would correspond to the 
free energy of the inflaton field
in the flat background limit. 

Subtraction schemes different from Eq.~\eqref{subt} can be 
proposed~\cite{inprep}.
Yet, a deeper analysis of the RG running of 
field-independent constant terms, and their extension for non-flat
backgrounds is beyond the scope of the present paper. An
investigation of the conditions that could fix, at least partially,
the subtraction scheme is 
currently in progress, and will be reported elsewhere~\cite{inprep}. 
In particular, a decisive 
question under investigation 
is whether or not the generation of the unphysical 
divergences for the constant, field-independent 
terms could be avoided via a modification 
of the conditions imposed on the regulator 
function~\eqref{regulator}.


\begin{thebibliography}{99}

\bibitem{inflation}
A. H. Guth, {\it Inflationary universe: A possible solution to the horizon and flatness problems}, Phys. Rev. D {\bf 23}, 347--356 (1981).

\bibitem{density-fluct}
A. A. Starobinsky, {\it A new type of isotropic cosmological models without singularity}, Phys. Lett. B {\bf 91}, 99--102 (1980);
V. F. Mukhanov and G. V. Chibisov, {\it Quantum fluctuations and a nonsingular universe}, JETP Lett. {\bf 33}, 532--535 (1981)
[Pisma Zh. Eksp. Teor. Fiz. {\bf 33}, 549--553 (1981)].

\bibitem{slow-roll}
A. D. Linde, {\it A new inflationary universe scenario: A possible solution of the horizon, flatness, homogeneity, isotropy and primordial monopole problems}, 
Phys. Lett. B {\bf 108}, 389--393 (1982);
A. Albrecht and P. J. Steinhardt, {\it Cosmology for Grand Unified Theories with Radiatively Induced Symmetry Breaking}, Phys. Rev. Lett. {\bf 48}, 1220--1223 (1982).

\bibitem{encyc}
J. Martin, C. Ringeval and V. Vennin, {\it Encyclopaedia Inflationaris}, Phys. Dark Univ. {\bf 5-6}, 75--235 (2014).

\bibitem{FLRW} 
A. Friedmann, {\it \" Uber die Kr\" ummung des Raumes}, Z. Phys. A {\bf 10}, 377--386 (1922);
G. Lema\^{\i}tre, {\it A Homogeneous Universe of Constant Mass and Increasing Radius accounting for the Radial Velocity of Extra-galactic Nebulae}, 
Mon. Not. R. Astron. Soc. {\bf 91}, 483--490 (1931);
H. P. Robertson, {\it Kinematics and World-Structure}, Astrophys. J. {\bf 82}, 284 (1935);
A. G. Walker, {\it On Milne's Theory of World---Structure}, Proc. London Math. Soc. {\bf 42}, 90--127 (1937).

\bibitem{planck}
P. A. R. Ade {\em et al.} [Planck collaboration], {\it Planck 2015 results. XIII. Cosmological parameters}, Astron. Astrophys. {\bf 594}, A13 (2016);
P. A. R. Ade {\em et al.} [Planck collaboration], {\it Planck 2015 results. XX. Constraints on inflation}, Astron. Astrophys. {\bf 594}, A20 (2016);
P. A. R. Ade {\em et al.} [BICEP2 and Keck Array collaboration], {\it {Improved Constraints on Cosmology and Foregrounds from BICEP2 and 
Keck Array Cosmic Microwave Background Data with Inclusion of 95 GHz Band}}, Phys. Rev. Lett. {\bf 116}, 031302 (2016).

\bibitem{chaotic}
A. D. Linde, {\it Chaotic inflation}, Phys. Lett B {\bf 129}, 177--181 (1983); {\it Eternally existing self-reproducing chaotic inflanationary universe}, 
Phys. Lett. B {\bf 175}, 395--400 (1986).  

\bibitem{criticism}
J. Earman and J. Mosterin, {\it A Critical Look at Inflationary Cosmology}, Philosophy of Science. {\bf 66}, 1--49 (1999);
P. J. Steinhardt, {\it The inflation debate}, Scientific American {\bf 304}, 36--43 (2011).

\bibitem{javier_higgs_inflation}
J. Rubio, {\it Higgs Inflation}, Front. Astron. Space Sci. {\bf 5}, 50 (2019).

\bibitem{rg_cosmo}
A. Bonanno and M. Reuter, {\it Cosmology of the Planck era from a renormalization group for quantum gravity}, Phys. Rev. D {\bf 65}, 043508 (2002);
A. Bonanno and M. Reuter, {\it Cosmology with self-adjusting vacuum energy density from a renormalization group fixed point}, Phys. Lett. B {\bf 527}, 9--17 (2002); 
A. Bonanno and M. Reuter, {\it Cosmological perturbations in renormalization group derived cosmologies}, Int. J. Mod. Phys. D {\bf 13}, 107--121 (2004); 
E. Bentivegna, A. Bonanno and M. Reuter, {\it Confronting the IR fixed point cosmology with high-redshift observations}, J. Cosmology Astropart. Phys.  {\bf 0401}, 001 (2004);
I. L. Shapiro, J. Sola, 
{\it The scaling evolution of cosmological constant},	
J. High Energy Phys. {\bf 2002}, 006--006 (2002);
I. L. Shapiro, J. Sola, 
{\it Massive fields temper anomaly-induced inflation: the clue to graceful exit?},
Phys. Lett. B {\bf 530}, 10--19 (2002);
I. L. Shapiro, J. Sola, 
{\it Scaling behavior of the cosmological constant and the possible existence of new forces and new light degrees of freedom},
Phys. Lett. B {\bf 475}, 236--246 (2000);
I. L. Shapiro, J. Sola, 
{\it Cosmological constant, renormalization group and Planck scale physics},
Nucl. Phys. B Proc. Suppl. {\bf 127}, 71--76 (2004);
I. L. Shapiro, J. Sola, H. Stefancic,
{\it Running G and $\Lambda$ at low energies from physics at $M_X$: possible cosmological and astrophysical implications},
J. Cosmol. Astropart. Phys. {\bf 2005}, 012--012 (2005);
I. L. Shapiro, J. Sola, 
{\it On the possible running of the cosmological ``constant"},
Phys. Lett. B {\bf 682}, 105--113 (2009).

\bibitem{rg_qeg}
Y.-F. Cai, Y.-C. Chang, P. Chen, D. A. Easson and T. Qiu, {\it Planck constraints on Higgs modulated reheating of renormalization group improved inflation}, 
Phys. Rev. D {\bf 88}, 083508 (2013);
G. Kofinas and V. Zarikas, {\it Asymptotically safe gravity and nonsingular inflationary big bang with vacuum birth}, Phys. Rev. D {\bf 94}, 103514 (2016);
R. Moti and A. Shoja, {\it On the effect of renormalization group improvement on the cosmological power spectrum}, Eur. Phys. J. C {\bf 78}, 32 (2018).

\bibitem{curved_RG1}
A. Kaya, {\it Exact renormalization group flow in an expanding Universe and screening of the cosmological constant}, Phys. Rev. D {\bf 87}, 123501 (2013).

\bibitem{curved_RG2}
J. Serreau, {\it Renormalization group flow and symmetry restoration in de Sitter space}, Phys. Lett. B {\bf 730}, 271--274 (2014); 
T. Prokopec and G. Rigopoulos, {\it Functional renormalization group for stochastic inflation}, J. Cosmology Astropart. Phys. {\bf 1808}, 013 (2018). 

\bibitem{curved_RG3}
M. Guilleux and J. Serreau, {\it Quantum scalar fields in de Sitter space from the nonperturbative renormalization group}, Phys. Rev. D {\bf 92}, 084010 (2015).

\bibitem{prd_kinvpropt}
B. Guberina, R. Horvat, and H. Stefancic, {\it Renormalization-group running of the cosmological constant and the fate of the universe}, Phys. Rev. D {\bf 67}, 083001 (2003);
M. Reuter and H. Weyer, {\it Renormalization group improved gravitational actions: A Brans-Dicke approach}, Phys. Rev. D {\bf 69}, 104022 (2004);
M. Reuter and H. Weyer, {\it Running Newton constant, improved gravitational actions, and galaxy rotation curves}, Phys. Rev. D {\bf 70}, 124028 (2004);
A. Babic, B. Guberina, R. Horvat, and H. Stefancic, {\it Renormalization-group running cosmologies: A scale-setting procedure}, Phys. Rev. D {\bf 71}, 124041  (2005);
A. Bonanno and M. Reuter, {\it Spacetime structure of an evaporating black hole in quantum gravity}, Phys. Rev. D {\bf 73}, 083005 (2006);
D. Lopez Nacir and F. D. Mazzitelli, {\it Running of Newton's constant and noninteger powers of the d'Alembertian}, Phys. Rev. D {\bf 75}, 024003 (2007);
A, Contillo, {\it Evolution of cosmological perturbations in a renormalization-group-driven inflationary scenario}, Phys. Rev. D {\bf 83}, 085016 (2011);
A. Contillo, M. Hindmarsh, and C. Rahmede, {\it Renormalization group improvement of scalar field inflation}, Phys. Rev. D {\bf 85}, 043501 (2012);
M. Hindmarsh and I. D. Saltas, {\it $f(R)$ gravity from the renormalization group}, Phys. Rev. D {\bf 86}, 064029 (2012); 
E. J. Copeland, C. Rahmede, and I. D. Saltas, {\it Asymptotically safe Starobinsky inflation}, Phys. Rev. D {\bf 91}, 103530 (2015);
G. D'Odorico and F. Saueressig, {\it Quantum phase transitions in the Belinsky-Khalatnikov-Lifshitz universe}, Phys. Rev. D {\bf 92}, 124068 (2015);
G. Kofinas and V. Zarikas, {\it Solution of the dark energy and its coincidence problem based on local antigravity sources without fine-tuning or new scales}, 
Phys. Rev. D {\bf 97}, 123542 (2018);
I. L. Shapiro, J. Sola, C. Espana-Bonet, 
P. Ruiz-Lapuente, {\it Variable Cosmological Constant 
as a Planck Scale Effect}, Phys. Lett. B {\bf 574}, 149--155 (2003);
C. Espana-Bonet, P. Ruiz-Lapuente, I. L. Shapiro, J. Sola, 
{\it Testing the running of the cosmological constant with Type Ia Supernovae at high z},
J. Cosmol. Astropart. Phys. {\bf 2004}, 006--006 (2004);
I. L. Shapiro, J. Sola, {\it A Friedmann-Lemaitre-Robertson-Walker cosmological model with running Lambda}, 
e-print astro-ph/0401015.

\bibitem{running_vacuum}
J. Sola, {\it Dark energy: a quantum fossil from the inflationary
Universe?}, J. Phys. A {\bf 41}, 164066 (2008);
J. Sola, A. Gomez-Valent, J. De Cruz Perez,
{\it First Evidence of Running Cosmic Vacuum: Challenging the Concordance Model},
Astrophys. J. {\bf 836}, 43 (2017);
S. Basilakos, N. E. Mavromatos, J. Sola Paracaula, 
{\it Quantum Anomalies in String-Inspired Running Vacuum Universe: 
Inflation and Axion Dark Matter}, 
Phys. Lett. B {\bf 803} 135342 (2020).

\bibitem{IJMPD}
J. Sola, A. Gomez-Valent,
{\it The $\overline{\Lambda}$CDM Cosmology: From inflation to dark energy
through running $\Lambda$},
Int. J. Mod. Phys. D {\bf 24}, 1541003 (2015).

\bibitem{PRD}
S. Basilakos, N. E. Mavromatos, J. Sola Paracaula, {\it Gravitational and 
Chiral Anomalies in the Running Vacuum Universe and 
Matter-Antimatter Asymmetry},
Phys. Rev. D {\bf 101}, 045001 (2020).

\bibitem{kadanoff}
L. P. Kadanoff, {\it Scaling laws for ising models near ${T}_{c}$}, Physics {\bf 2}, 263--272 (1966).

\bibitem{wilson}
K. G. Wilson, {\it Model Hamiltonians for Local Quantum Field Theory}, Phys. Rev. {\bf 140}, B445--B457 (1965)

\bibitem{polonyi}
J. Polonyi, {\it Lectures on the functional renormalization group method}, Cent. Eur. J. Phys. {\bf 1}, 1--71 (2003).

\bibitem{baumann}
D. Baumann, {\it TASI Lectures on Inflation}, available as
arXiv:0907.5424 [hep-th].

\bibitem{nat_infl}
K. Freese, J. A. Frieman and A. V. Olinto, {\it Natural inflation with pseudo Nambu-Goldstone bosons}, Phys. Rev. Lett. {\bf 65}, 3233--3236 (1990);
K. Freese and W. H. Kinney, {\it Natural inflation: consistency with cosmic microwave background observations of Planck and BICEP2}, 
J. Cosmology Astropart. Phys. {\bf 1503}, 044 (2015);
P. Creminelli, D. Lopez Nacir, M. Simonovic, G. Trevisan, and M. Zaldarriaga, {\it ${\ensuremath{\phi}}^{2}$ or Not ${\ensuremath{\phi}}^{2}$: 
Testing the Simplest Inflationary Potential}, Phys. Rev. Lett. {\bf 112}, 241303 (2014).

\bibitem{linear_periodic}
T. Kobayashi, O. Seto and Y. Yamaguchi, {\it Axion monodromy inflation with sinusoidal corrections}, Prog. Theor. Exp. Phys. {\bf 2014}, 103E01 (2014);
T. Higaki, T. Kobayashi, O. Seto and Y. Yamaguchi, {\it Axion monodromy inflation with multi-natural modulations}, J. Cosmology Astropart. Phys.  {\bf 1410}, 025 (2014).

\bibitem{eea_rg}
C. Wetterich, {\it Exact evolution equation for the effective potential}, Phys. Lett. B {\bf 301}, 90--94 (1993);  
T. R. Morris, {\it The Exact renormalization group and approximate solutions}, Int. J. Mod. Phys. A {\bf 9}, 2411--2449 (1994).

\bibitem{sk}
L. V. Keldysh, {\it Diagram technique for nonequilibrium processes}, Zh. Eksp. Teor. Fiz. {\bf 47}, 1515--1527 (1964);  [Sov. Phys. JETP {\bf 20}, 1018 (1965)];
O. V. Konstantinov and V. I. Perel, {\it A graphical technique for computation of kinetic quantities}, Zh. Eksp. Teor. Fiz. {\bf 39}, 197 (1960);  [Sov. Phys. JETP {\bf 12}, 142 (1961)];
J. Schwinger, {\it Brownian motion of a quantum oscillator}, J. Math. Phys. {\bf 2}, 407--432 (1961);
L. P.  Kadanoff and G. Baym, {\it Quantum Statistical Mechanics}, (Benjamin, New York) (1962).

\bibitem{sk_new}
J. Polonyi, {\it Quantum-classical crossover in electrodynamics}, Phys. Rev. D {\bf 74}, 065014 (2006);
S. Nagy, J. Polonyi, I. Steib, {\it Quantum renormalization group}, Phys. Rev. D {\bf 93}, 025008 (2016);
S. Nagy, J. Polonyi, I. Steib, {\it Euclidean scalar field theory in the bilocal approximation}, Phys. Rev. D {\bf 97}, 085002 (2018).

\bibitem{overshoot_prob}
R. Brustein, S. P. de Alwis and P. Martens, {\it Cosmological stabilization of moduli with steep potentials}, Phys. Rev. D {\bf 70}, 126012 (2004). 

\bibitem{Codello2015}
A. Codello, N. Defenu and G. D'Odorico, {\it Critical exponents of $O(N)$ models in fractional dimension}, Phys. Rev. D {\bf 91}, 105003 (2015). 

\bibitem{Defenu2018}
N. Defenu and A. Codello, {\it Scaling solutions in the derivative expansion}, Phys. Rev. D {\bf 98}, 016013 (2018).

\bibitem{Litim2000}
D. F. Litim, {\it Optimisation of the exact renormalisation group}, Phys. Lett. B {\bf 486}, 92--99 (2000).

\bibitem{msg_lpa}
I. Nandori, S. Nagy, K. Sailer and A. Trombettoni, {\it Comparison of renormalization group schemes for sine-Gordon-type models}, Phys. Rev. D {\bf 80}, 025008 (2009); 
I. Nandori, {\it On the renormalization of the bosonized multi-flavor Schwinger model}, Phys. Lett. B {\bf 662}, 302--308 (2008);
S. Nagy, I. Nandori, J. Polonyi and K. Sailer, {\it Generalized universality in the massive sine-Gordon model}, Phys. Rev. D {\bf 77}, 025026 (2008). 

\bibitem{msg_beyond_lpa}
I. Nandori, {\it Bosonization and functional renormalization group approach in the framework of ${\mathrm{QED}}_{2}$}, Phys. Rev. D {\bf 84}, 065024 (2011).

\bibitem{Gi2006}
H. Gies, 
{\it Introduction to the Functional RG and Applications to Gauge Theories}, 
in {\em Springer Notes in Physics vol.~852}, 
J. Polonyi and A.~Schwenk (Eds.), 
pp.~287--348 Springer (Heidelberg), 2012.
See also e-print hep-ph/0611146.

\bibitem{De2007}
B. Delamotte, {\it An Introduction to the Nonperturbative Renormalization Group}, 
in {\em Springer Notes in Physics vol.~852}, 
J. Polonyi and A.~Schwenk (Eds.), 
pp.~49--85, Springer (Heidelberg), 2012.
See also e-print cond-mat/0702365.
\color{black}

\bibitem{lyth_2}
D. H. Lyth, {\it Particle Physics Models of Inflation}, Lect. Notes Phys. {\bf 738} 81-118 (2008).

\bibitem{lyth_1}
D. H. Lyth and A. Riotto, {\it Particle physics models of inflation and the cosmological density perturbation}, Phys. Rept. {\bf 314} 1-146 (1999).

\bibitem{pathintbook}
R. J. Rivers, {\it Path Integral Methods in Quantum Field Theory}, 
(Cambridge University Press, 1987).

\bibitem{inprep}
U. D. Jentschura, N. Defenu, I. G. M\'ari\'an, I. N\'andori,
and A. Trombettoni,
{\em Role of Field-Independent Terms in Nonperturbative RG},
in preparation (2019).

\end{thebibliography}
\end{document}